\documentclass[pre,showpacs]{revtex4}
\usepackage{graphicx}
\usepackage{epsfig}
\usepackage{amsmath}
\usepackage{amssymb}
\usepackage{subfigure}
\begin{document}

\title{Sliding periodic boundary conditions for lattice Boltzmann and lattice kinetic equations}

\author{R. Adhikari, J.-C. Desplat, K. Stratford}
\affiliation{School of Physics, JCMB Kings Buildings, Mayfield Road, Edinburgh
EH9 3JZ GB}

\begin{abstract} We present a method to impose linear shear flow in discrete-velocity kinetic models of hydrodynamics through the use of sliding periodic boundary conditions. Our method is derived by an explicit coarse-graining of the Lees-Edwards boundary conditions for Couette flow in molecular dynamics, followed by a projection of the resulting equations onto the subspace spanned by the discrete velocities of the lattice Boltzmann method. The boundary conditions are obtained without resort to perturbative expansions or modifications of the discrete velocity equilibria, allowing our method to be applied to a wide class of lattice Boltzmann models. Our numerical results for the sheared hydrodynamics of a one-component isothermal fluid show excellent agreement with analytical results, while for a two-component fluid the results show a clear improvement over previous methods for introducing Lees-Edwards boundary conditions into lattice Boltzmann. Using our method, we obtain a dynamical steady state in a sheared spinodally decomposing two-dimensional fluid, under conditions where previous methods give spurious finite-size artifacts. 
\end{abstract}
\maketitle

\section{INTRODUCTION} Planar Couette flow, a volume-preserving flow with a
constant velocity gradient, is frequently used to probe the response of soft
matter to shear \cite{Cates:2000asi}.  Soft materials display a wide range of
interesting behaviour in shear flow: shear suppresses the critical temperature
in a critical fluid \cite{Onuki:1997}; destroys long-range order in certain
phases \cite{Ackerson:1981} while inducing it in others \cite{Cates:1989i2l}; and can
lead to complex rheological behaviour in liquid-crystalline phases
\cite{Bandyopadhyay:2000}.

Couette flow is realised in the laboratory by confining fluid between two concentric cylinders which are then set into relative motion. Forces are transmitted
from the cylinders to the fluid, setting up flows adjacent to the plates, which
finally develop to give a linear velocity profile between them. The profile deviates from linearity in a narrow region near the
plates called the Knudsen layer, where the fluid velocity changes rapidly. In
the majority of cases, the width of the Knudsen layer is negligible compared to
the separation between the plates and the velocity profile may be assumed to be
linear throughout. The resulting rheological response is that of the bulk and
has negligible contributions from boundary effects. When the radius of the cylinders is large compared to their separation, the flow approximates that between two parallel plates, a geometry known as planar Couette flow. 

A direct replication of such a setup in a computer simulation provides less
satisfactory results: the width of the boundary layer or of the inhomogeneities
introduced by the walls is often comparable to the size of the central layer where
the flow profile is linear. A simulation where bulk behaviour is desired is
then largely contaminated with boundary effects. Reducing such boundary effects
by increasing the overall size of the simulation is inefficient, and when
computational power is limited, often impossible.  

In simulations without shear flow, boundary effects are eliminated by the use
of the Born-von Karman periodic boundary conditions (PBC). In 1972 Lees and
Edwards \cite{Lees:1972} proposed a remarkable algorithm which consists of a
simple modification of PBC that enables the simulation of Couette flow without
the introduction of \emph{physical} boundaries. The flow has no boundary layer
and no inhomogeneities are introduced. The algorithm is extremely useful when
steady-state bulk behaviour needs to be probed in an accurate and efficient way \cite{Evans:1990book}. For reasons explained below, we shall refer to the algorithm of
Lees and Edwards as sliding periodic boundary conditions (SPBC).

The SPBC was originally formulated for use in molecular dynamics. However,
Couette flow can be described in at least two other frameworks, those of
kinetic theory and hydrodynamics. Molecular dynamics, kinetic theory and
hydrodynamics form a hierarchy of descriptions, in which one can proceed
from the most detailed level (molecular dynamics) to the least detailed
(hydrodynamics) by a ``coarse-graining'' procedure. The Bogoliubov \cite{Liboff:1990book} method of
passing from Newton's equation of motion to the Boltzmann equation and the
Chapman-Enskog expansion \cite{Chapman:1991book}, used in passing from kinetic theory to hydrodynamics, are examples of such "coarse-graining" procedures. It is reasonable, then, to
expect that algorithms developed for molecular dynamics may find use in kinetic
theory, and the resulting kinetic method may be useful for the simulation of
hydrodynamic flow. In this paper we show that SPBC in molecular dynamics can be
\emph{exactly} represented in terms of a scattering kernel in kinetic theory,
and that this kinetic boundary condition can be used in the lattice Boltzmann
equation to provide an useful method for the simulation of the hydrodynamics of
linear shear flow which reproduces bulk behaviour accurately and efficiently.

There have been two earlier efforts at eliminating boundary effects in LBE with shear flow. Wagner and Yeomans \cite{Wagner:1999} used modified equilibria to impose a change in momentum. Subsequently, Wagner and Pagonabarraga (WP)\cite{Wagner:2002} proposed a heuristic transformation rule (which required no change in the LBE equilibria) on the one-particle distributions, based on a perturbative expansion of the distributions around the shear flow steady state. The truncation of the perturbative expansion implicit in their work leads to accumulative numerical error at large times. Such errors are avoided by the algorithm presented in this work.

The plan of the paper is as follows. In the following section we provide a general summary of kinetic methods like the LBE and introduce a new kinetic scheme for the Cahn-Hilliard equation. We combine these two methods to produce a LBM for binary fluids. In Sec.\ref{boundary}, which contains the main results of this paper, we derive the kinetic analogue of the Lees-Edwards SPBC in the continuum using the formalism of kinetic boundary conditions. The difficulties associated with the discretisation of these boundary conditions on the lattice are discussed and solved. In Sec.\ref{results} we present detailed numerical results to show that our algorithm compares favourably with the earlier work of Wagner and Pagonabarraga. We conclude with a summary of our results and directions for future work.

\section{DISCRETE VELOCITY KINETIC MODELS}\label{DVKM}
The recent resurgence of interest in discrete kinetic models \cite{chen:98, succi-book:01} has
mainly been due to the discovery that a specially discretised Boltzmann
equation can be used as a flexible Navier-Stokes solver. This lattice Boltzmann
equation (LBE) was originally motivated by the desire to improve lattice gas
models \cite{McNamara:88, higuera:89}, but was later recognised to be a special velocity discretisation of the Boltzmann equation \cite{Shan:1998,He:1997b}. The LBE represents a \emph{hyperbolic} set
of partial differential equations (PDEs) with a constant, linear advection
operator whereas the Navier-Stokes equations are \emph{parabolic} with a
time-dependent non-linear advection operator. An accurate numerical solution of
the former is easier than that of the latter. The LBE numerical algorithm
is explicit and involves nearest-neighbours only, making it heavily
parallelisable. Finally, complicated boundary conditions can easily be
incorporated with little additional effort. Due to this combination of
numerical and computational advantage, the LBE is now widely considered to be
competitive with more traditional methods of computational fluid dynamics \cite{succi-book:01}. 

The LBE has spawned a large number of variants which collectively go under the name of the lattice Boltzmann method (LBM). They share the basic philosophy of using a set of hyperbolic PDEs of the form
\begin{equation}\label{Ker}
\partial_t f_i +{\bf c}_i\cdot\nabla f_i = - {\cal L}_{ij}(f_j - f_j^0)
\end{equation}
(often abbreviated as kinetic equations) to solve a set of parabolic PDEs in the form of conservation laws
\begin{equation}
\partial_t C^{a} + \nabla j^{a} = 0
\end{equation}
for the set of \emph{conserved} variables $C^{a}$. The connection between the two sets of equations is established by defining the conserved variables to be velocity moments of the ``distribution'' functions $f_i$ appearing in Eq.\ref{Ker}. The matrix ${\cal L}_{ij}$ depends on certain symmetry requirements, while the ``equilibrium'' distribution functions $f_i^0$ depend on the fluxes $j^{a}$ which,  in general, are non-linear functions of the conserved variables $C^{a}$. Parabolic conservation laws arise in a variety of physical situations where diffusion is accompanied by advection. The incompressible Navier-Stokes equations representing the diffusion of vorticity and its advection by the flow field, and the Cahn-Hilliard equation describing the dynamics of a conserved scalar, are both of this type. 

In certain instances, the kinetic equations Eq.\ref{Ker} follow from an underlying physically meaningful kinetic description where the continuum analogue of the ``equilibrium'' distribution function $f_i^0$ describes true statistical mechanical equilibrium. This is true in the LBE, where the continuum analogue of $f_i^0$ is the Maxwell-Boltzmann distribution. In other instances, the kinetic equations do not  follow from Boltzmann kinetic theory and the ``equilibrium'' distribution functions do not have the same statistical mechanical significance as in the LBE. In such circumstances, the LBM is best viewed as a kinetic relaxational scheme \cite{Perthame:2002book} with the $f_i^0$ dictating the fixed point of the relaxation process and attempts at a statistical interpretation of the kinetic equations are best avoided. To keep this distinction in mind, we shall refer to relaxation schemes whose equilibria do not follow from a Boltzmann equation as lattice kinetic equations (LKE) and reserve the terminology of LBE for kinetic equations derived from Boltzmann equilibria. This distinction is important when assessing the accuracy of the SPBC for the LBE and LKE methods. Below we describe in detail the LBM for binary fluids, which combines the LBE for solving the Navier-Stokes equations with a new LKE method for the Cahn-Hilliard equation which is closely related to LBM of Swift {\em et al} \cite{Swift:96}.
 
\subsection{LBE for isothermal Navier-Stokes}
The set of conservation laws of mass and momentum $\partial_t \rho + \nabla\cdot{\bf g} = 0$ and $\partial_t {\bf g} + \nabla\cdot\Pi = 0$ contain the hydrodynamic description of an isothermal fluid. They correspond to the dynamics of two conserved quantities, the mass density $\rho$, and the momentum density ${\bf g}$. The momentum density is itself the mass flux ${\bf g}=\rho{\bf v}$, while $\Pi_{\alpha\beta}$ is the momentum flux. The fluid velocity is defined by ${\bf v}={\bf g}/\rho$. The momentum flux for a \emph{incompressible} Newtonian fluid is given by
\begin{equation}
\Pi_{\alpha\beta} = g_{\alpha}v_{\beta} + p\delta_{\alpha\beta} - \eta(\nabla_{\alpha}
v_{\beta} + \nabla_{\beta}v_{\alpha})
\end{equation}
Together with the conservation laws, this lead to the Navier-Stokes equations of isothermal fluid flow
\begin{equation}
\rho(\partial_t{\bf v} + {\bf v\cdot\nabla v}) = -\nabla p + \eta\nabla^{2}{\bf
v}
\end{equation}
The lattice Boltzmann equation (LBE) is derived from the discrete velocity
BGK equation
\begin{equation}
\label{dbe}
\partial_t f_i +{\bf c}_i\cdot\nabla f_i = - {\cal L}_{ij}(f_j - f_j^0)
\end{equation}
by a further discretisation of space and time. The discrete velocities $\{{\bf
c}_i\}$ are the nodes of a Gauss-Hermite quadrature \cite{Shan:1998, He:1997b} which
ensure that the conserved moments of the distribution functions have the same
values as in the continuum. For the isothermal LBE, these are the mass and
momentum densities,
\begin{equation}\label{defmoments}
\sum_i f_i\{1, c_{i\alpha}\} = \{\rho,\rho v_{\alpha}\}
\end{equation}
with Greek indices denoting Cartesian directions. The Boltzmann kinetic
description is restricted to a dilute gas with
an ideal equation of state $p = nk_BT = \rho c_s^2$, where $n$ is the
number density and $c_s$ is the isothermal sound speed. It is then
convenient to subtract this trivial kinetic contribution to the pressure
from the momentum flux tensor and define a \emph{deviatoric} momentum
flux 
\begin{equation}\label{contstress}
S_{\alpha\beta} = \Pi_{\alpha\beta} - \rho c_s^2\delta_{\alpha\beta} = \sum_i
f_i Q_{i\alpha\beta}
\end{equation}
where $Q_{i\alpha\beta} = c_{i\alpha}c_{i\beta}-c_s^2\delta_{\alpha\beta}$ and
$\Pi_{\alpha\beta}=\sum_i f_i c_{i\alpha}c_{i\beta}$ is the momentum flux. In
the most commonly used D$d$Q$n$ models\cite{Koelman:91, Qian:92}containing $n$ quadrature nodes in $d$ dimensions, the equilibrium distribution functions are given by
\begin{equation}
f_i^0 = w_i\left[\rho + {\rho v_{\alpha}c_{i\alpha}\over c_s^2}
+ {\rho v_{\alpha}v_{\beta}Q_{i\alpha\beta}\over 2c_s^4}\right]
\end{equation}
where $w_i$ are the set of weights defining the quadrature and repeated Greek
indices are summed over. This form is obtained by retaining the first three terms in the  Hermite polynomial expansion of the Maxwell-Boltzmann distribution \cite{Shan:1998}.
Finally, the BGK collision matrix ${\cal L}_{ij}$ must satisfy the constraints
imposed by the conservation laws and rotational symmetry. In the single
time approximation ${\cal L}_{ij} = \tau^{-1}\delta_{ij}$, the shear ($\eta)$
and bulk viscosities ($\zeta$) are related to the relaxation time $\tau$ by
$\eta = \rho c_s^2\tau$, $\zeta = (2/d)\rho c_s^2\tau$. For multiple relaxation
time models, the above relations remain valid respectively, if $\tau$ is replaced by the relaxation time of the shear and bulk viscous stress \cite{Ladd:2001}.

To derive a numerical scheme, the set of coupled hyperbolic equations in Eq.
\ref{dbe} must be integrated. We integrate the discrete BGK equation along a
characteristic \cite{He:1998nonideal, Dellar:2003} for a time $\Delta t$ to obtain,
\begin{equation}
\label{lbe1}
f_i({\bf x} + {\bf c}_i\Delta t, t+\Delta t) -  f_i({\bf x},t) = 
\int_{0}^{\Delta t}ds~ J_i({\bf x}+{\bf c}_is, t+s)
\end{equation}
where $J_i({\bf x},{\bf c}_i,t) = -{\cal L}_{ij}(f_j-f_j^0)$. The integral above
may be approximated to second-order accuracy using the trapezium rule to give
\begin{equation}
\label{lbetrapz}
f_i({\bf x} + {\bf c}_i\Delta t, t+\Delta t) -  f_i({\bf x},t) = 
{\Delta t\over2}\left[ J_i({\bf x}+{\bf c}_i\Delta t, t+\Delta t)
+ J_i({\bf x}, t)\right]
\end{equation}
This represents a set of implicit equations which would normally be
solved by matrix inversion. However, the transformation to a new set
of distribution functions $f_i^{'}$ \cite{Dellar:2003} defined by
\begin{eqnarray}
\label{transformf1}
f^{'}_i({\bf x},t) = f_i({\bf x},t) - {\Delta t\over2}J_i({\bf x},t) 
\end{eqnarray}
renders the system explicit. Solving the above for the $f_i$ in terms of the
$f_i^{'}$ gives
\begin{equation}
\label{transformf2}
f_i - f_i^0 = ({\mathbf 1}+\frac{\Delta t}{2}{\cal 
L})^{-1}_{ij}(f_j^{'}-f_j^0)
\end{equation}
Now combining Eq.\ref{lbetrapz},\ref{transformf1},and \ref{transformf2}, we
obtain the second-order accurate \emph{explicit} LBE scheme as
\begin{equation}
\label{mrtlbe}
f_i^{'}({\bf x}+{\bf c}_i\Delta t,t+\Delta t) = f_i^{'}({\bf x},t) -\Delta t {\cal
L}_{ik}({\mathbf 1} + \frac{\Delta t}{2}{\cal 
L})^{-1}_{kj}(f_j^{'}-f_j^0) 
\end{equation}
In the single time approximation ${\cal L}_{ij}=\delta_{ij}/\tau$, the above reduces to the familiar LBGK form 
\begin{equation}
\label{mrtlbe}
f_i^{'}({\bf x}+{\bf c}_i\Delta t,t+\Delta t) = f_i^{'}({\bf x},t)
\end{equation}
Eq.\ref{transformf1} allows us to \emph{calculate} the hydrodynamic moments
using the new distribution functions in terms of their earlier \emph{definitions},
Eq.\ref{defmoments}. Taking the zeroth and first moments of Eq.\ref{transformf1} and using the constraints of mass and momentum conservation $\sum_i J_i = \sum_i J_i
c_{i\alpha} = 0 $, we find that mass and momentum densities are calculated as
before, with $f_i$ replaced by $f_i^{\prime}$. However, the second moment $\sum_i J_i Q_{i\alpha\beta}\propto (S_{\alpha\beta}-\rho v_{\alpha}v_{\beta}$) is not constrained to be vanish, implying that that deviatoric stress must be calculated using
\begin{equation}\label{discretestress}
S_{\alpha\beta} =\left[\sum_i f_i^{\prime}Q_{i\alpha\beta}+{\Delta t\over2\tau} \rho v_{\alpha}v_{\beta}\right]/(1+ {\Delta t\over2\tau})
\end{equation}
Since the conserved moments $\rho$ and $\rho v_{\alpha}$ which determine the equilibrium distributions $f_i^0$ can be calculated using either the $f_i$ or the $f_i^{\prime}$, the $f_i$ are redundant and the numerical scheme can be formulated entirely in terms of the $f_i^{\prime}$. Notice that $f_i^{'}\rightarrow f_i$ as $\Delta t \rightarrow 0$, implying that $f_i^{'}$ encodes discrete lattice effects. This is true, in particular, for the formula for the deviatoric stress (Eq.\ref{discretestress})  which goes over to the continuum formula (Eq.\ref{contstress}) as $\Delta t/\tau\rightarrow 0$.

Mathematically, the passage to the hydrodynamic limit can be demonstrated using the Chapman-Enskog multiple scale expansion \cite{chen:98} or Grad's method of moment closure \cite{succi-book:01}. Physically, hydrodynamic behaviour is obtained when the mean free path of the gas is small compared to the length scales associated with the flow. In the Navier-Stokes regime, hydrodynamic relaxation takes place predominantly through the diffusion of vorticity rather than the propagation of sound. For a flow with typical flow velocity $v$, a length scale $L$, and a viscosity $\eta$, the time scales associated with sound propagation and vorticity diffusion are  $L/c_s$ and $\rho L^2/\eta$ respectively. This ratio, which we define to be the Knudsen number, can be conveniently expressed as $\hbox{\it Kn} = \hbox{\it Ma}/\hbox{\it Re}$, where $\hbox{\it Ma} = v/c_s$ and $\hbox{\it Re} = \rho vL/\eta$ are the Mach and Reynolds numbers. The LBE reproduces incompressible Navier-Stokes hydrodynamics {\em only} for $\hbox{\it Kn}\ll1$ and $\hbox{\it Ma}\ll1$.
%

\subsection{LKE for Cahn-Hilliard}

The Cahn-Hilliard equation (also known as Model $B$ in the Hohenberg-Halperin classification of critical dynamics \cite{Chaikin:2000}), 
\begin{equation}
\label{che}
\partial_t \psi + \nabla\cdot(\psi{\bf v} -M\nabla\mu) = 0
\end{equation}
corresponds to the conservation law of a single conserved variable $\psi$ with a flux $\psi{\bf v} -M\nabla\mu$. The chemical potential $\mu$ is obtained from the Landau-Ginzburg-Wilson functional $F[\psi]$ controlling the equilibrium fluctuations in $\psi$ through $\mu = \delta F[\psi]/\delta\psi$. The mobility $M$ is assumed to be constant and independent of $\psi$, though the velocity ${\bf v}$ is in general both space and time dependent. A kinetic relaxation scheme for the above is obtained by introducing  distributions $g_i$ obeying the kinetic relaxation equation
\begin{equation}
\label{kre}
\partial_t g_i +{\bf c}_i\cdot\nabla g_i = - {\cal L}^{\psi}_{ij}(g_j - g_j^0)
\end{equation}
with $\psi = \sum_i g_i$, and $j_{\alpha} = \sum_i g_i c_{i\alpha}$. The model is fully determined by specifying the collision matrix ${\cal L}^{\psi}$, the stationary function $g_i^0$, and the nodes of the quadrature $\{{\bf c}_i\}$. In what follows we used the standard D$d$Q$n$ quadrature of the LBE, a single-time collision matrix, ${\cal L}^{\psi}_{ij}=\delta_{ij}/\tau_{\psi}$ and the following stationary distributions
\begin{equation}\label{newgi0}
g_i^0 = w_i\left[{\psi v_{\alpha}c_{i\alpha}\over c_s^2} + 
{(\psi v_{\alpha}v_{\beta} + \mu\delta_{\alpha\beta})Q_{i\alpha\beta}\over 2c_s^4}\right] + \delta_{i0}\psi
\end{equation}
where $\delta_{i0} = 1$ for the zero-velocity population but zero otherwise. The motivation for this choice is provided elsewhere \cite{Stratford:2005}, but note that it satisfies the following constraints \cite{Swift:96} on its first three moments: $\sum_i g_i^0 = \psi$, $\sum_i g_i^0 c_{i\alpha} = \psi v_{\alpha}$, and $\sum_i g_i^0 c_{i\alpha}c_{i\beta} = \mu\delta_{\alpha\beta} + \psi v_{\alpha}v_{\beta}$.
A fully discrete second-order numerical scheme is obtained by the characteristics-based integration method introduced in the previous section. As before, this leads to an implicit set of equations for the $g_i$, which can be rendered explicit by introducing the new distributions $g_i^{\prime}$
\begin{equation}
g_i^{\prime} = g_i - {\Delta t\over2}J_i^{\psi}
\end{equation}
where $J_i^{\psi} = -{\cal L}_{ij}^{\psi}(g_i - g_i^0)$. This gives the fully discrete LKE
\begin{equation}
g_i^{'}({\bf x}+{\bf c}_i\Delta t,t+\Delta t) = g_i^{'}({\bf x},t)
-(g_i^{'}-g_i^0)/(\tau_{\psi} + {\Delta t\over2}) 
\end{equation}
Using the constraint of order parameter conservation $\sum_i J_i^{\psi} = 0$ we find that the order parameter is given as in the continuum by $\psi = \sum_i g_i^{\prime}$, but the flux has discrete lattice corrections and must be calculated using
\begin{equation}
j_{\alpha} = \left[\sum_i g_i^{\prime}c_{i\alpha} + {\Delta t\over2\tau_{\psi}}\psi v_{\alpha}\right]/(1 + {\Delta t\over 2\tau_{\psi}})
\end{equation}
As with the deviatoric stress in Navier-Stokes (Eq.\ref{discretestress}) the flux formula reduces to its continuum definition as $\Delta t/\tau_{\psi}\rightarrow 0$. The velocity and chemical potential in the stationary distribution $g_i^0$ are prescribed and so do not need to be determined through the moments of $g_i$. 

The formulation presented above is closely related to the binary fluid LBM of Swift \emph{et al} \cite{Swift:96}. Our model differs from theirs in two important ways. First, our model is formulated in the continuum from which we derive a second-order accurate discrete model using the method of characteristics; Swift \emph{et al} begin with a discrete model which is only first-order accurate. Second, our stationary distributions $g_i^0$ are distinct from their ``equilibrium'' distributions, though both give identical moments up to the second. We find that with our choice of $g_i^0$ both isotropy and numerical stability is greatly enhanced. A detailed comparison of the two methods is presented elsewhere \cite{Stratford:2005}; here we only present comparative results for shear flow using SPBC.

\subsection{LBM for binary fluids}
The equations of binary fluid hydrodynamics result from combining the Cahn-Hilliard equation with the Navier-Stokes equation with an additional force density representing the exchange of momentum between the two components of the fluid. We present them for completeness:
\begin{equation}
\rho(\partial_t{\bf v} + {\bf v\cdot\nabla v}) = -\nabla p + \eta\nabla^{2}{\bf v} + \psi\nabla\mu
\end{equation}
\begin{equation}
\partial_t \psi + \nabla\cdot(\psi{\bf v} -M\nabla\mu) = 0
\end{equation}
The order parameter $\psi$ now represents the local concentration, whose equilibrium fluctuations are governed by $F[\psi] = \int d^3{\bf x}[A\psi^2 + B\psi^4 +\kappa(\nabla\psi)^2]$ \cite{kendon:2001jfm}. The additional $\psi\nabla\mu$ term in the momentum equation can be written as the divergence of a stress tensor $\sigma_{\alpha\beta}^{\psi}$, such that $\nabla_{\beta}\sigma_{\alpha\beta}^{\psi} = \psi\nabla_{\alpha}\mu$. Then, there are at least two ways of incorporating this information into the LBE. The first, followed by Swift \emph{et al} simply modifies the definition of the equilibrium deviatoric stress $S_{\alpha\beta}^{0}$\\\ to include this term. Thus the equilibrium distributions for the $f_i$ are now given by
\begin{equation}
f_i^0 = w_i\left[\rho + {\rho v_{\alpha}c_{i\alpha}\over c_s^2}
+ {(\rho v_{\alpha}v_{\beta}+\sigma_{\alpha\beta}^{\psi})Q_{i\alpha\beta}
\over 2c_s^4}\right]
\end{equation}
The above ``equilibrium'' distribution can no longer be derived using Gauss-Hermite quadrature of a stationary solution of an underlying Boltzmann equation and is best thought of as an \emph{ansatz} to be justified post-facto. Further, this definition breaks Galilean invariance and is responsible for the some of the artifacts seen in the model \cite{kendon:2001jfm},\cite{Luo:2000}. An alternative and more attractive method \cite{Adhikari:2005} is to treat the $\psi\nabla\mu$ term as an \emph{external} force density in the forced Boltzmann equation. This requires no change to the LBE equilibria and is Galilean invariant by construction. In this paper we follow the method of Swift \emph{et al} so that a like-for-like comparison can be made with earlier shear flow algorithms \cite{Wagner:2002}. No change is required in the LKE for Cahn-Hilliard equation, other than identifying the velocity $\bf v$ with that in the Navier-Stokes equation. This completes our presentation of the LBM for binary fluids. In the next section we show how SPBC is implemented in the model which can then be used to study the hydrodynamics of a sheared binary fluid.

\section{BOUNDARY CONDITIONS FOR PLANAR COUETTE FLOW}
\label{boundary}

\subsection{Molecular dynamics} 
Let us briefly recall the SPBC proposed by Lees and Edwards \cite{Lees:1972}, assuming the simulation unit cell to be a parallelepiped with faces at $x=\pm L$, $y=\pm L$ and $z=\pm L$. Periodic boundary conditions provide a prescription to deal with coordinate values that flow out of this range as the integration of the equations of motion proceeds. PBC simply identifies all coordinates modulo the size of the unit cell $2L$: any coordinate which lies outside the range $[-L,L]$ is mapped
back into it using modular  division. Additionally, the PBC is applied to each
coordinate separately and does not affect the velocities of the particles.
Thus, with PBC applied in the $y$ direction, a particle leaving the top face at
$y=L$ is simply reinserted at the bottom face at $y=-L$, with no change to the
$y$ and $z$ coordinates or the velocity. Similar rules apply for PBC in the $x$
and $z$ directions. 

The SPBC proposed by Lees and Edwards follows from a simple modification of
PBC. It prescribes that a particle leaving a face of the box in the
velocity-gradient direction is re-entered at the opposite face but at a new
position displaced along the flow direction, and with a new velocity
incremented only in the flow direction. The ordinary PBC is applied for particles
leaving the box in the remaining directions.  For concreteness, assume that the
flow and velocity-gradient directions are $x$ and $y$ respectively. The flow
velocities at $\pm y = L$ are $\pm U$, giving a velocity gradient $\dot\gamma =
U/L$. The SPBC then requires that a particle leaving the top face at $y=L$  with
coordinates $(x, L, z)$ and velocities $(c_x,c_y,c_z)$ is reinserted at the
bottom face $y=-L$ with a new $x$-coordinate $x^{\prime}$ and new $x$-component of
velocity $c_x^{\prime}$ given by 
\begin{equation}\label{SPBC}
x^{\prime} = x - 2Ut,~~ c_x^{\prime} = c_x - 2U
\end{equation}
so  that its final coordinates and velocities are $(x^{\prime}, -L, z)$ and
$(c_x^{\prime}, c_y, c_z)$ respectively. Particles leaving the box through the
bottom face at $y=-L$  re-enter at the top face at $y=L$ using the same
rule as in Eq.\ref{SPBC} but with the sign of $U$ reversed. The standard PBC is applied to particles leaving the box through the faces at $x=\pm L$ and $z = \pm L$. 

The motivation for these rules becomes transparent when we consider the
deformation of a cube with edges of length $L$ in uniform shear flow with
$\dot\gamma = U/L$. The cube undergoes an affine deformation where the top and
bottom faces move in opposite directions with relative velocity $2U$. Points at
the top and bottom faces which had identical $x$-coordinates at $t = 0$ are
displaced by $2Ut$ at time $t$.

\subsection{Kinetic theory} 
The Lees-Edwards sliding periodic boundary conditions described above operate
on the coordinates and velocities of the particles crossing the boundary
$\partial V$ of the simulation volume $V$. The kinetic description of a
molecular system, on the other hand, is formulated not in terms of particle
coordinates and velocities, but in terms of the one-particle distribution
function $f({\bf x},{\bf c},t)$, representing the number density of particles
at the phase point $({\bf x},{\bf c})$ at time $t$. Accordingly, to make use of
SPBC in kinetic theory, we must translate the SPBC condition Eq.\ref{SPBC} into
an \emph{equivalent} condition on the one-particle distribution function. This
can be done using the scattering kernel approach for reflective boundary
conditions in kinetic theory \cite{Cercignani:1988book}, as we show below. 

Consider, as before, fluid flow in a domain $V$ whose boundary is $\partial V$.
Let ${\bf s}\in\partial V$ be a point on the boundary and ${\bf n}({\bf s})$ be
the inward pointing normal at that point. Molecules at the phase point $({\bf
s}, {\bf c})$ leave $V$ and impinge on the boundary if ${\bf n}({\bf
s})\cdot{\bf c}<0$; similarly, molecules enter $V$ and emerge from the boundary
if ${\bf n}({\bf s})\cdot{\bf c}>0$. Let $B({\bf s},{\bf c}\rightarrow{\bf
s^{\prime}},{\bf c^{\prime}}; t)$ be the probability that at time $t$, a
molecule which strikes the boundary at point ${\bf s}$ with velocity ${\bf c}$
emerges at point ${\bf s^{\prime}}$ with velocity ${\bf c}^{\prime}$. For
a non-absorbing wall, a molecule is always re-emitted, which provides the
normalisation condition for $P$
\begin{equation}
\int d{\bf s} d{\bf c} B({\bf s},{\bf c}\rightarrow{\bf s^{\prime}},{\bf
c^{\prime}}; t)\Theta(-{\bf c}\cdot{\bf n}) = 1
\end{equation}
The total number of molecules emerging from the boundary (per unit time and area) at the point ${\bf s^{\prime}}$ with velocity ${\bf c^{\prime}}$ must be equal to the number of molecules of all velocities incident on all parts of the boundary that change their velocity to ${\bf c^{\prime}}$ and emerge at the point ${\bf s^{\prime}}$. Stated mathematically \cite{Cercignani:1988book},
\begin{equation}\label{kbc1}
\Theta({\bf c^{\prime}}\cdot{\bf n^{\prime}})|{\bf c^{\prime}}\cdot{\bf n^{\prime}}|f({\bf s^{\prime}},{\bf c^{\prime}},t) = 
\int d{\bf s} d{\bf c} B({\bf s},{\bf c}\rightarrow{\bf s^{\prime}},{\bf
c^{\prime}}; t)
\Theta(-{\bf c}\cdot{\bf n})|{\bf c}\cdot{\bf
n}|f({\bf s},{\bf c},t).
\end{equation}
we obtain the boundary condition on the populations emerging from the boundary
in terms of the incident populations and the scattering kernel $B$ which
characterises the particle-wall interaction. The Heaviside step function
$\Theta$ selects the impinging and emerging populations from the total
populations at the boundary. The majority of physical boundary conditions are
local in space and independent of time, yielding a scattering kernel $B$ that
is dependent non-trivially on the velocities only. In the simplest case of
specular reflection, the particle velocity normal to the wall is reversed at
the point of contact. In the opposite case of diffuse scattering, the particle
is thermalised on contact and leaves the wall with a velocity sampled from a
Maxwellian distribution determined by the wall temperature.
%
%
The scattering kernel for specular scattering is
$B({\bf s},{\bf c}\rightarrow{\bf s}^{\prime},{\bf c^{\prime}}) =
\delta({\bf s}-{\bf s}^{\prime})\delta({\bf c} + 2{\bf n}({\bf n}\cdot{\bf
c})-{\bf c}^{\prime})$
while that for diffuse scattering is
$B({\bf s},{\bf c}\rightarrow{\bf s}^{\prime},{\bf c^{\prime}}) =
\delta({\bf s}-{\bf s}^{\prime})|{\bf n}\cdot{\bf
c}|\exp(-c^2/k_BT)/2\pi(k_BT)^2$
More realistic wall-particle interactions are obviously possible and have been
explored in detail \cite{Cercignani:1988book} and implemented in the LBE \cite{Ansumali:2002}.

Viewed in this light, Lees-Edwards boundary conditions can be thought of as a {\em non-local} scattering process, in which particles incident on the upper boundary cross over to be re-emitted from the lower boundary, with the initial and final
coordinates and momenta being given by Eq.\ref{SPBC}. It is clear, then, that the transformation rules for the particle coordinates and velocities at the boundary can be used to construct the scattering kernel $B$. Enforcing the SPBC rules (Eq.\ref{SPBC}) at the boundaries through delta functions, we find that
\begin{equation} B_{SPBC}({\bf x},{\bf c}\rightarrow{\bf x^{\prime}},{\bf c^{\prime}};t)
=\delta(x \pm2Ut - x^{\prime})\delta(y + y^{\prime})\delta(z - z^{\prime})
\delta(c_x \pm 2U- c_x^{\prime})\delta(c_y - c_y^{\prime})\delta(c_z -
c_z^{\prime}) \end{equation}
Inserting this into the Eq.\ref{kbc1} and completing the integration, we obtain
the equivalent of Lees-Edwards boundary conditions for the distribution function:
\begin{equation}\label{SPBC-KE}
f(x,y,z,c_x,c_y,c_z,t) = f(x\pm2Ut,-y,z,c_x\pm2U, c_y,c_z,t)
\end{equation}
Below we shall implement this SPBC on the distribution functions
of the LBE and LKE, paying special attention to the difficulties in going from 
a space where ${\bf x},{\bf c}$ are continuous to a discrete velocity space and a discrete space-time lattice.

\subsection{Discrete implementation}
The SPBC, Eq.\ref{SPBC-KE}, poses no problem so far as the temporal discretisation is concerned, since it relates values of the distribution function at equal times. However, a direct implementation of Eq.\ref{SPBC-KE} on the lattice is complicated by the fact that the nodes of the quadrature (ie. the set of velocities ${\bf c}_i$) are discrete, as is the position ${\bf x}$. The discreteness of the velocity space must be dealt with even if the model is formulated in the continuum, as is the discrete velocity BGK equation Eq.\ref{dbe}. For such a model to be computationally useful, it must ultimately live on the lattice, requiring us to address the discretisation of space as well. Thus there are two distinct, and conceptually separate problems: we deal with the velocity discretisation first. Let us first recall that any (well-behaved) function can in general be expanded in terms of orthogonal polynomials. A particularly useful expansion for the $f$ is in terms of Hermite polynomials \cite{Grad:1949}, where the coefficients of the expansions are Hermite moments of the distribution function. This expansion is the basis of the derivation of the discrete velocity BGK equation from the Boltzmann-BGK equation \cite{Shan:1998}. For an isothermal LBE, at least three terms must be retained in this expansion. The expansion coefficients are evaluated by Gaussian quadrature \cite{He:1997b}, the nodes of the quadrature defining the discrete velocities ${\bf c}_i$. The $f_i({\bf x},t)$ can then be represented in terms of the velocity moments of $f({\bf x},{\bf c},t)$ as
\begin{equation}\label{fiexp}
f_i = w_i\left[\rho + {\rho v_{\alpha}c_{i\alpha}\over c_s^2}
+ {S_{\alpha\beta}Q_{i\alpha\beta}\over 2c_s^4}+ G_i \right]
\end{equation}
where $G_i$ represent higher moments of the distribution function, variously known as kinetic modes, non-hydrodynamic modes, or ``ghost'' modes \cite{benzi:92, Dellar:2002,dHumieres:1992}. We circumvent the problem of the discreteness of velocity space as follows: instead of implementing the SPBC Eq.\ref{SPBC-KE} directly in terms of the distributions $f({\bf x},{\bf c}, t)$, we obtain equivalent conditions on their moments, which are then ``projected'' on the discrete velocity space by using Eq.\ref{fiexp}. Taking the first three moments of Eq.\ref{SPBC-KE} followed by a change of variables in the integrals we obtain the SPBC conditions on the mass, momentum and stress densities as: 
\begin{eqnarray}
\rho &=&\rho^{\prime} \\
\rho v_{\alpha} &=& (\rho v_{\alpha})^{\prime}\pm \rho^{\prime}U_{\alpha}\\  S_{\alpha\beta} &=& S^{\prime}_{\alpha\beta} \pm (\rho v_{\alpha})^{\prime}U_{\beta} \pm (\rho v_{\beta})^{\prime}U_{\alpha} + \rho^{\prime}U_{\alpha}U_{\beta}
\end{eqnarray}
where the $\rho = \rho(x,y,z,t)$, $\rho^{\prime} = \rho(x\pm Ut,\mp y,z,t)$ etc, and $U_{\alpha} = \delta_{x\alpha}U$. We now demand that the first three moments of the Eq.\ref{SPBC-KE} be satisfied at the discrete level. This gives us the discrete analogue of the SPBC for the LBE.
\begin{equation}
f_i(\rho,\rho v_{\alpha}, S_{\alpha\beta}, G) = f_i^{\prime}(\rho^{\prime}, (\rho v_{\alpha})^{\prime},S^{\prime}_{\alpha\beta},G)
\end{equation}
The transformation of the model dependent kinetic modes $G_i$ have been ignored for clarity of presentation, but can easily be included in the above framework.(We list all the modes for the D$d$Q$15$ model in \cite{Adhikari:2005}; for D$d$Q$9$ see \cite{Dellar:2002}). Below we discuss why neglecting the transformation of the kinetic modes may be justified.

The SPBC for the LKE for the Cahn-Hilliard equation can be implemented by repeating the steps above. First we note that the $g_i$ can be written as
\begin{equation}
g_i = w_i\left[{j_{\alpha}c_{i\alpha}\over c_s^2} + {\mu_{\alpha\beta}Q_{i\alpha\beta}\over 2 c_s^4} + G_i^{\psi}\right] + \delta_{i0}\psi
\end{equation}
where $\mu_{\alpha\beta} = \sum_ig_iQ_{i\alpha\beta}$, and $G_i^{\psi}$ represent the kinetic modes. Since the transformation rule for the moments of the distribution function is determined by completely by the number of velocities that enter into its definition, the transformation rules for the LKE moments follow identically from the counterparts in the LBE by making the replacements $\rho\rightarrow\psi$, $(\rho v_{\alpha})\rightarrow j_{\alpha}$, and $S_{\alpha\beta}\rightarrow\mu_{\alpha\beta}$. We thus have:
\begin{eqnarray}
\psi &=&\psi^{\prime} \\
j{\alpha} &=&j_{\alpha}^{\prime}\pm \psi^{\prime}U_{\alpha}\\  
\mu_{\alpha\beta} &=& \mu^{\prime}_{\alpha\beta} \pm j_{\alpha}^{\prime}U_{\beta} \pm j_{\beta}^{\prime}U_{\alpha} + \psi^{\prime}U_{\alpha}U_{\beta}
\end{eqnarray}
Leaving the kinetic modes untransformed, we obtain the discrete SPBC for the LKE as
\begin{equation}
g_i(\psi,j_{\alpha}, \mu_{\alpha\beta}, \Gamma^{\psi}) = g_i^{\prime}(\psi^{\prime}, j_{\alpha}^{\prime},\mu^{\prime}_{\alpha\beta},\Gamma^{\psi})
\end{equation}
A linear interpolation is applied to deal with the case when $x - Ut$ is not a member of the lattice. The distribution function at the point $x - Ut$ is partitioned between the nodes bracketing the point $x - Ut$ in proportion to their distance from the point. With the transformation rules and the linear interpolation we now have a complete discrete SPBC algorithm. Notice that we arrive at the algorithm from the continuum SPBC using two approximations: first, we do not transform the kinetic modes and second, we apply a linear interpolation to deal with the spatial discreteness. The first of the approximations is easily avoided, once expressions for kinetic modes are known in terms of the microscopic velocities. In the hydrodynamic limit, the kinetic modes are only weakly coupled to the conserved variables \cite{Adhikari:2004} and their transformation may be neglected to a first approximation. In the interests of simplicity, we have only presented the model-independent transformation rules for the hydrodynamic modes. The second of the approximations is unavoidable on a discrete lattice, though more accurate interpolation schemes might be envisaged. As we show in the following section, the present combination already gives results of acceptable numerical accuracy.

\subsection{Large shear rates}
The LBE constraint of small Mach numbers ($U\ll c_s$) places an upper limit on the shear rate $\dot\gamma = U/L$ which becomes progressively smaller with increasing system size $L$. To overcome this limitation, Wagner and Pagonabarraga \cite{Wagner:1999} introduced internal Lees-Edwards boundaries, where the Lees-Edwards boundary conditions are applied \emph{without} the application of PBC. In the particle picture, this corresponds to particles crossing the boundary accompanied by a change in velocity $\pm U$ and a change in the $x-$ coordinate $\pm Ut$, but no change in the $y$ and $z$ coordinates. In terms of distribution functions, we have 
\begin{equation}\label{LE-KE}
f(x,y,z,c_x,c_y,c_z,t) = f(x\pm2Ut,y,z,c_x\pm2U, c_y,c_z,t)
\end{equation}
whose discrete implementation follows as in the previous section, except that PBC is no longer required in the $y-$ direction. With the introduction of $N_{LE}$ such planes, the maximum attainable shear rate is $N_{LE}$ times that with the SPBC. In the following section we demonstrate the use of LE planes in the study of spinodal decomposition under shear.
 
\section{Test applications}\label{results}
We now provide results of numerical simulations using the SPBC derived in the previous section, treating in turn the hydrodynamics of sheared one-component and two-component fluids. In what follows, we define lattice units of time, length and mass by setting $\Delta t = 1$, $\Delta x=1$, $\rho_0 = 1$, where $\rho_0$ is the \emph{mean} mass density at a lattice node. In our simulations, we use a collision operator which relaxes the stress modes at a rate $\tau^{-1}$ and projects out all the non-hydrodynamic modes \cite{Behrend:1994}, \cite{Ladd:2001}. For a simple fluid the only adjustable parameter, then, is the relaxation time $\tau$, now related to the shear viscosity by $\eta = \rho_0 c_s^2\tau$. For a binary fluid, the adjustable parameters are the mobility $M$, the interfacial thickness $\xi$ and the interfacial surface tension $\sigma$. In the LKE the mobility is related to $\tau_{\psi}$ by 
$M=c_s^2\tau_{\psi}$, while the parameters contained in the LGW functional $F[\psi]$ determine $\xi$ and $\sigma$ \cite{kendon:2001jfm}. Our choice for all the binary fluid parameters is that of Kendon {\emph et al} \cite{kendon:2001jfm}. In all cases, the simulation is performed in a rectangular domain of dimension $L_x\times L_y$. Shear is applied so that flow takes place along the $x-$axis and the flow gradient is along the $y-$axis.

\subsection{Single fluid}
\begin{figure}
\centering
\subfigure[$\eta$ = 1.41]
{
\includegraphics[]{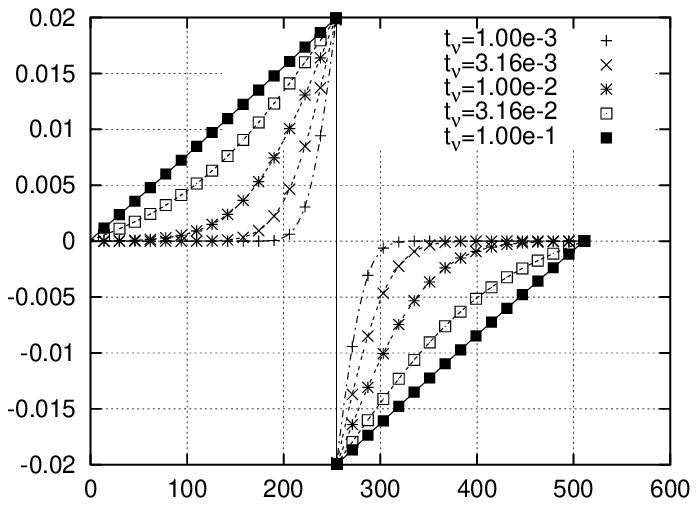}
\label{startup1}
}
\subfigure[$\eta$ = 0.20]
{
\includegraphics[]{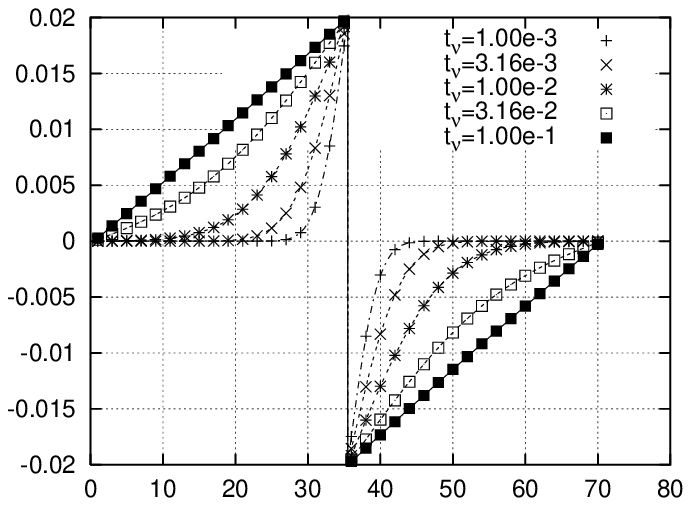}
\label{startup2}
} 
\subfigure[$\eta$ = 0.0005]
{
\includegraphics[]{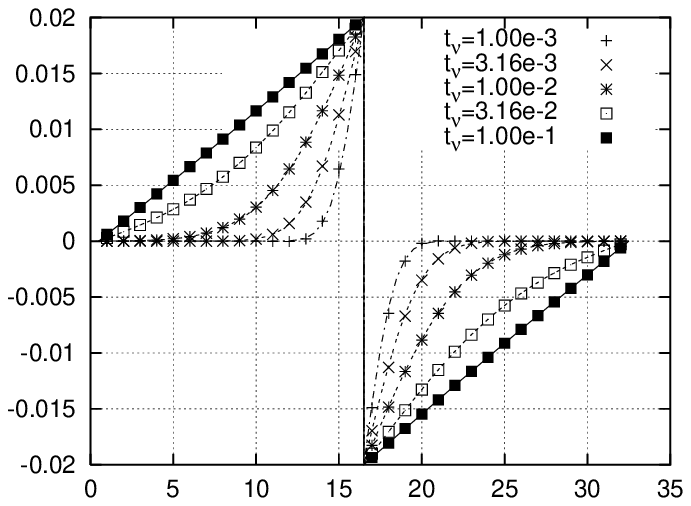}
\label{startup3}
}
\caption{Relaxation to steady-state in impulsively started planar Couette flow using the SPBC algorithm. The horizontal axis represents the $y$ axis for the lattice, the vertical axis corresponds to the flow velocity $v_{x}(y) = \sum_{x}v_{x}(x,y)/L_x$. Curves within each set correspond to different dimensionless times $t_{\nu}=\eta t/L_y^2 $. The symbols represent numerical data, dotted lines the theoretical expression.}
\end{figure}

For the single fluid we validate our method by comparing against two cases where analytical solutions are known. The first of these is impulsively started Couette flow, where a quiescent fluid is suddenly subjected to shearing motion. The relaxation of the velocity to its asymptotic linear shear profile is determined by the diffusion of momentum from the boundary into the bulk: the solution for flow between parallel plates moving at relative velocity $U$ with separation $L$ is \cite{Batchelor:1967}
\begin{equation}\label{couette}
v_x(x,y,z,t) = U(1-y/L) -{2U\over\pi}\sum_{n=1}^{\infty}\exp\left(-n^2\pi^2{\nu t\over L^2}\right)\sin{n\pi y\over L}
\end{equation}
where the kinematic viscosity $\nu=\eta/\rho_0$  is identical to $\eta$ in our choice of units. The SPBC algorithm can be used to study transient phenomena if the time scale of the transient is large compared to the time it takes for vorticity to diffuse across the transverse dimension of the system. To compare this with the results of an LBE simulation we must ensure that that the Knudsen number $Kn = \nu/L c_s$ is small, so that the the vorticity diffusion time scale $L^2/\nu$ is much larger than the sound propagation time scale $L/c_s$, and incompressible hydrodynamic behaviour is obtained. Three representative results from a wide range of viscosities shown in Figs~\ref{startup1}, \ref{startup2} and \ref{startup3}, where the numerical results are compared with the theoretical expression in Eq.\ref{couette}. Table~\ref{tab:sf1-all-relerr-LEMRT} shows the relative error $\epsilon_{v} = (v_{n} - v_{t})/v_{t}$, where $v_{n}$ is the numerical results, and $v_{t}$ the theoretical expression, half a 
lattice site away from the LE plane. As can be seen, the error
increases substantially with decreasing viscosity and decreases with time as the velocity profile relaxes to its steady state. Overall, errors remain very low in most situations apart from the very early stages of the most inertial runs. For this test, we find very little difference between our algorithm and that of the Wagner and Pagonabarraga for the range of parameters we have have explored; accordingly we do not display data for their algorithm.

Our second test explores the effect of a Gallilean transformation along the direction of the velocity gradient, or equivalently, that of an initial condition in which the fluid is moving uniformly with a velocity $U^{\perp}$ in the direction of the velocity gradient. In this case, we expect that the momentum of the system in the \emph{flow} direction will increase \emph{linearly} with time. This is most clearly understood in terms of SPBC for molecular dynamics: the boundary condition does work on each particle as it crosses the boundary, since it changes the velocity of particle in the flow direction by $\pm U$. A useful description is in terms of Lees-Edwards ``demons'' sitting at the boundary and giving a transverse impulse to every particle crossing the boundary. Since the total number of particles in the simulation box is constant, the number of particles leaving from the boundary is exactly balanced by the number of particles entering through the boundary. This implies that summed over the incoming and outgoing particles, the demons add equal but opposite amounts of momentum to the system, resulting in no change of momentum in the flow direction. (However, the kinetic energy, being proportional to $U^2$ does not cancel; the demons add energy to the system which is dissipated as heat.) However, with a net transverse flow, there is a net flux of particle crossing the boundary in a given direction and the momentum added by the demons no longer cancels out, resulting in an increase (or decrease) of momentum along the flow direction. For a constant transverse flux, the same amount of momentum is added at every time step, resulting in a linear increase of the transverse momentum. The same argument applies in kinetic theory. In the LBE, the energy density is not conserved, so the effects of heating are not noticeable. However, the argument for the increase in momentum still holds. Formalising our previous argument we see that the total momentum should increase as
\begin{equation}\label{LEdemon}
\sum_{{\bf x},i} f_i^{'}({\bf x},t)c_{ix} = \sum_{\bf x}\rho v_{x} = \rho_0 U^{\perp}L_{x}Ut
\end{equation}
which is simply the product of the  number of particle crossing the boundary $\rho_0U^{\perp}L_x$, the momentum added per particle $U$, and the duration of this process $t$. This should lead to a steady drift with time of the entire velocity profile.

\begin{figure}
\centering
\subfigure[]
{
\includegraphics[]{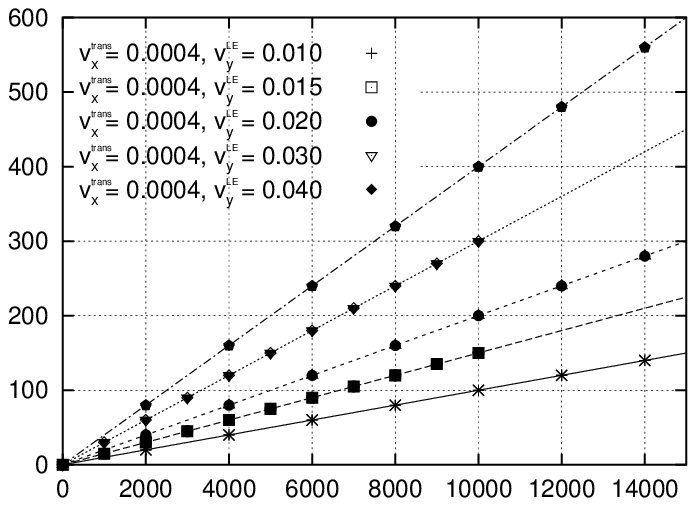}
\label{demon1}
}
\subfigure[]
{
\includegraphics[]{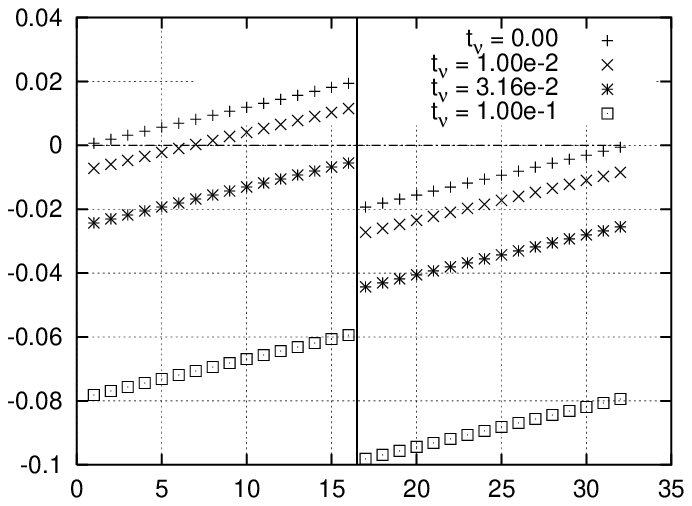}
\label{demon2}
}
\caption{Gallilean invariance of the SPBC algorithm. (a)Variation of the total fluid momentum in the flow direction as a function of time with flow in the velocity-gradient direction. Data is shown for $U = \{0.01;0.02;0.04\}$ and $U =\{0.015;0.03\}$). The vertical axis corresponds to $\sum_{x}(\rho v_{x})/\rho_0 U^{\perp}L_y$. Each line corresponds to the predicted values obtained
using eqn~\ref{couette}, where each line corresponds to runs
with identical $U$. Note that symbols are
overlap for any given $U$ and time. (b) Velocity profile $v_y = f(x)$ with 
$U=0.04$ and $U^{\perp}=0.004$. The horizontal axis represents the flow direction, the vertical axis represents the average velocity component $v_x$ 
at a location $y$. Simulation data has been plotted for $4t_{\nu}$}
\end{figure}

Our results for this test are shown in Fig.\ref{demon1}. The system was initialised with a linear velocity profile with a transverse velocity $U_{\perp}$ and all three components of the total momentum was measured. As expected, there was no change in the $y-$ or $z-$components, while the $x-$component increased linearly with time with a slope predicted by Eq.\ref{LEdemon}. Further, the entire velocity profile shifted by at a constant rate as shown in Fig.\ref{demon2}. This validates the  Gallilean invariance of the method. Similar simulations performed with the algorithm of Wagner and Pagonabarraga show no increase (the $x-$momentum remains constant) which clearly demonstrates the lack of Gallilean invariance. This has been noted in previous work where a sheared binary fluid simulation with a net transverse momentum produced unphysical results \cite{Cates:2004philtrans}. We have verified that mass was properly conserved in both our method and that of Wagner and Pagonabarraga.
\subsection{Binary fluids}
For the binary fluid, we have performed tests for steady-state deformation of a droplet in shear flow and spinodal decomposition under shear. In either case, to the best of our knowledge, no analytical results are available in the range of parameters accessible in a LBM simulation. Therefore, we must satisfy ourselves with qualitative tests based on a visual inspection of the data, focusing on possible artifacts near lattice planes where SPBC is applied. In the binary fluid case a like-for-like comparison with WP requires care since our binary fluid LBM differs significantly from theirs, even before the different handling of SPBC is introduced. Our preliminary investigations indicate that the $g_i^0$ used by WP introduce the largest source of error over and above those that may be introduced by their boundary conditions. Accordingly, we present results for spinodal decomposition and droplet deformation for the WP algorithm using the LBM outlined in their original work (henceforth WP0), the WP algorithm using the stationary distributions in Eq.\ref{newgi0} (henceforth WP1) and our LBM algorithm with SPBC as presented in Sec.\ref{DVKM} (henceforth SPBC).

\begin{figure} 
\subfigure[]
{
\includegraphics[width = 0.4\textwidth]{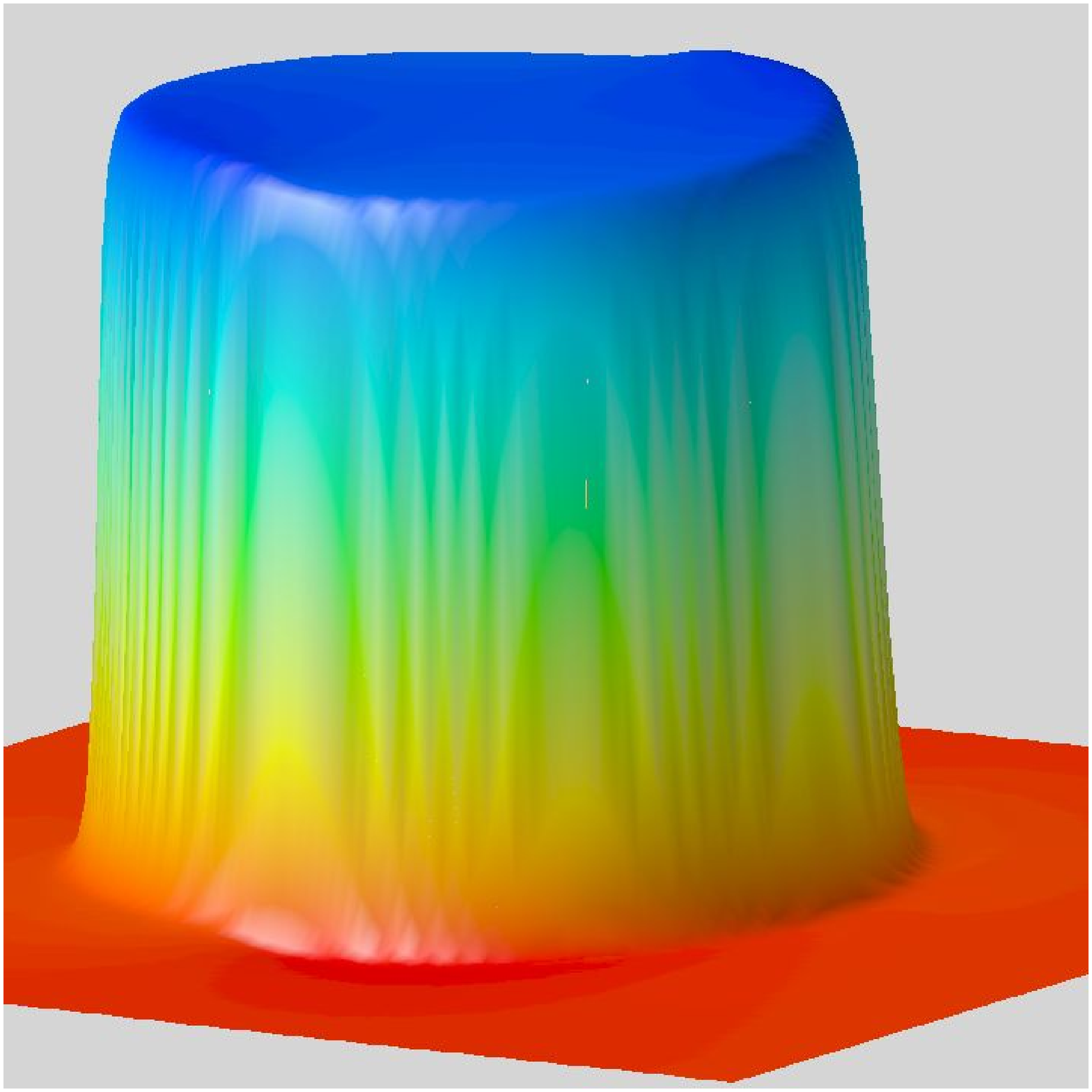}
}
\subfigure[]
{
\includegraphics[width = 0.4\textwidth]{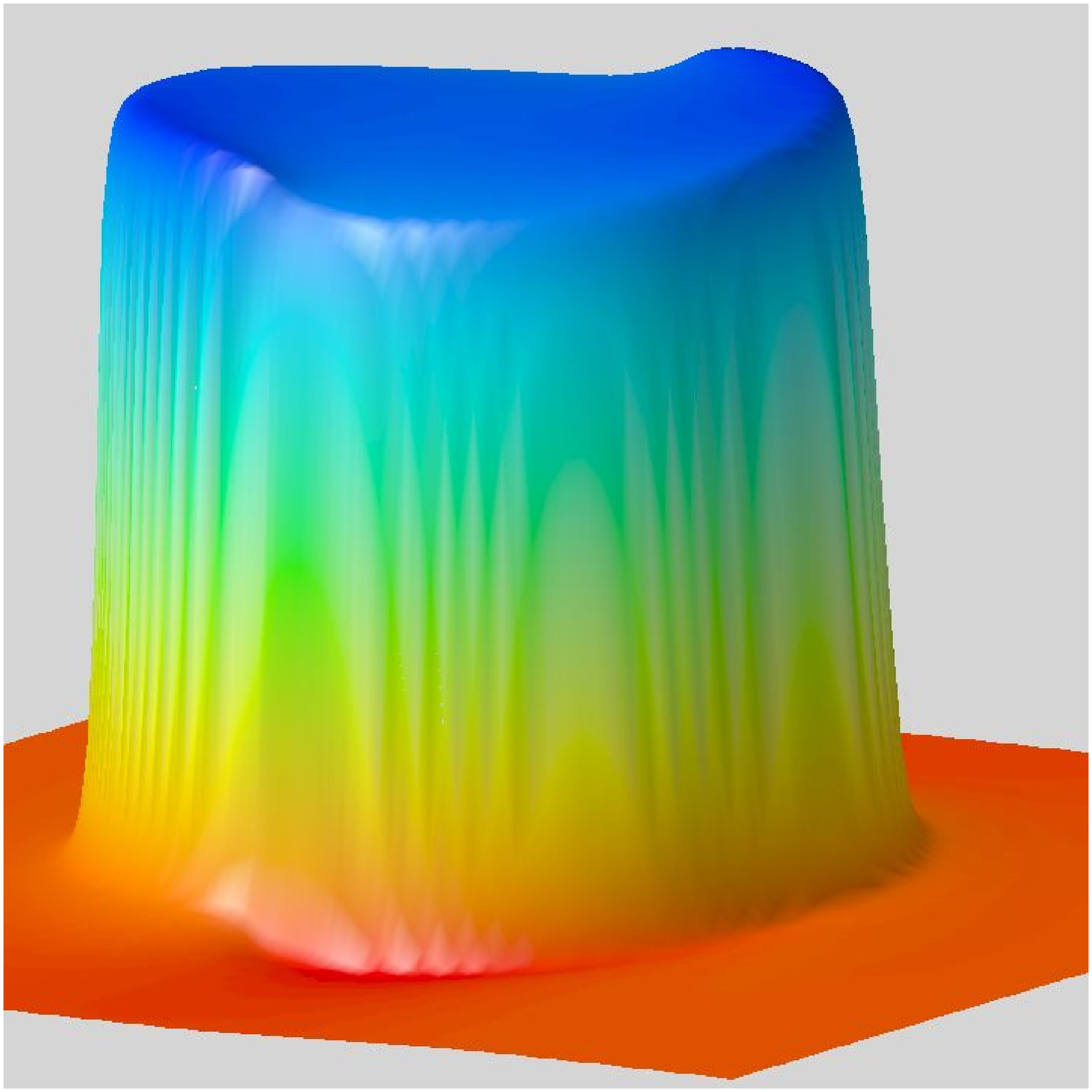} 
}
\caption{Surface plot $z =\psi(x,y)$ of the steady-state order concentration $\psi$ under shear at t = 100,000 for (a) SPBC and (b) WP0\label{drop:all3D}. Notice the sharp discontinuity in the order parameter field at the SPBC plane in (b).}
\end{figure}

\begin{figure} 
\subfigure[]
{
\includegraphics[width = 0.5\textwidth]{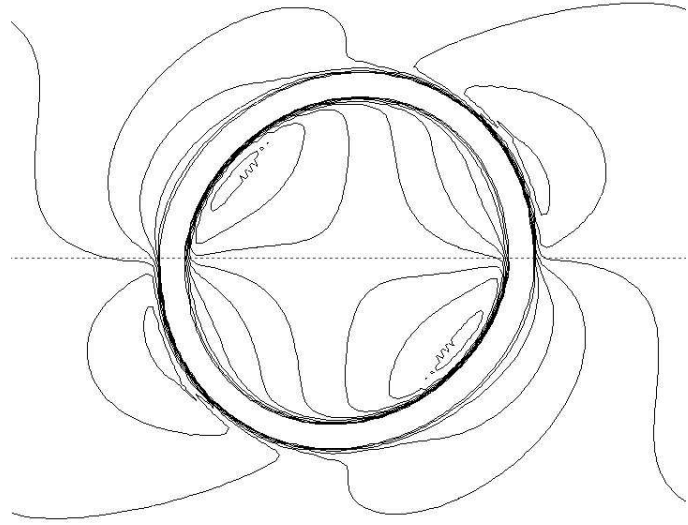}
}
\subfigure[]
{
\includegraphics[width = 0.5\textwidth]{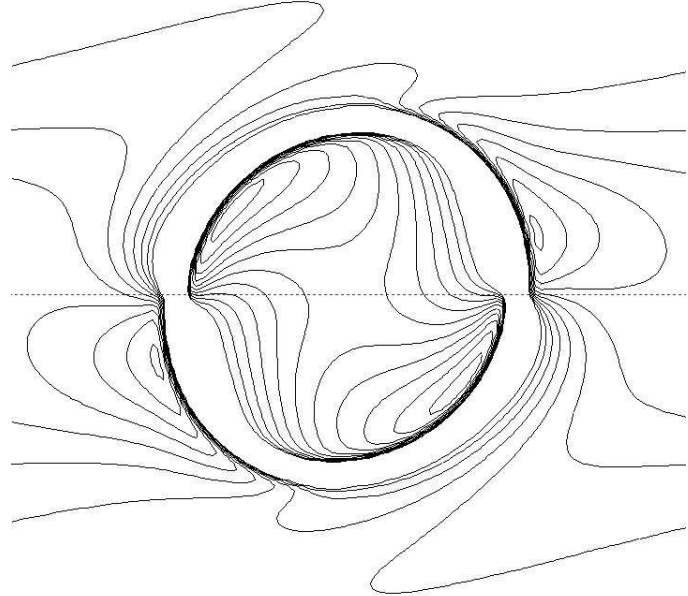} 
}
\caption{Steady state droplet deformation at t = 100,000 for (a) SPBC and
(b) WP0. Isolines start for $\Phi = -0.95$ with steps $-0.01$
inside the droplet, and for $\Phi = 0.95$ with steps $0.01$ outside. Artifacts at the SPBC plane indicated by the dotted line are clearly visible in the (b). \label{drop:all2Diso}}
\end{figure}

For our first test we looked at the steady-state deformation of a sheared droplet comparing SPBC with WP0. For a droplet of radius $R$ and surface tension $\sigma$ in a shear flow with shear rate $\dot\gamma$ the important dimensionless groups are the Reynolds number at the scale of the droplet $Re = \rho\dot\gamma R^2/\eta$ and the capillary number $Ca = \eta\dot\gamma R/\sigma$. The capillary number is the ratio of the viscous stress to the surface tension.  The system was sheared using one SPBC plane with $U = 0.032$ for system of size $128 \times 128$, resulting in a shear rate of $\dot{\gamma} = 2.50 \times 10^{-4}$. In the initial state a droplet of radius $R=32$ was centred respect to the SPBC boundary. The fluid viscosity was chosen such that $Ca = 0.038$  and Reynolds number $Re = 1.28$ with their ratio being $Re / Ca = 33.6$. The droplet was then visualised in the steady-state ($t = 100000$). Fig.~\ref{drop:all3D} clearly shows the existence of artifacts where the LE plane crosses the droplet. In particular, one notices the
presence of a cusp in the visualisation in both WP0 and SPBC schemes, but this artifact significantly more marked with WP0 than with our new scheme. In the isoline plot of Fig.~\ref{drop:all2Diso}, spurious behaviour is clearly observed at the boundary in the  WP0 scheme, but is almost unobservable SPBC. 

\begin{figure} 
\subfigure[]
{
\includegraphics[width = 0.4\textwidth]{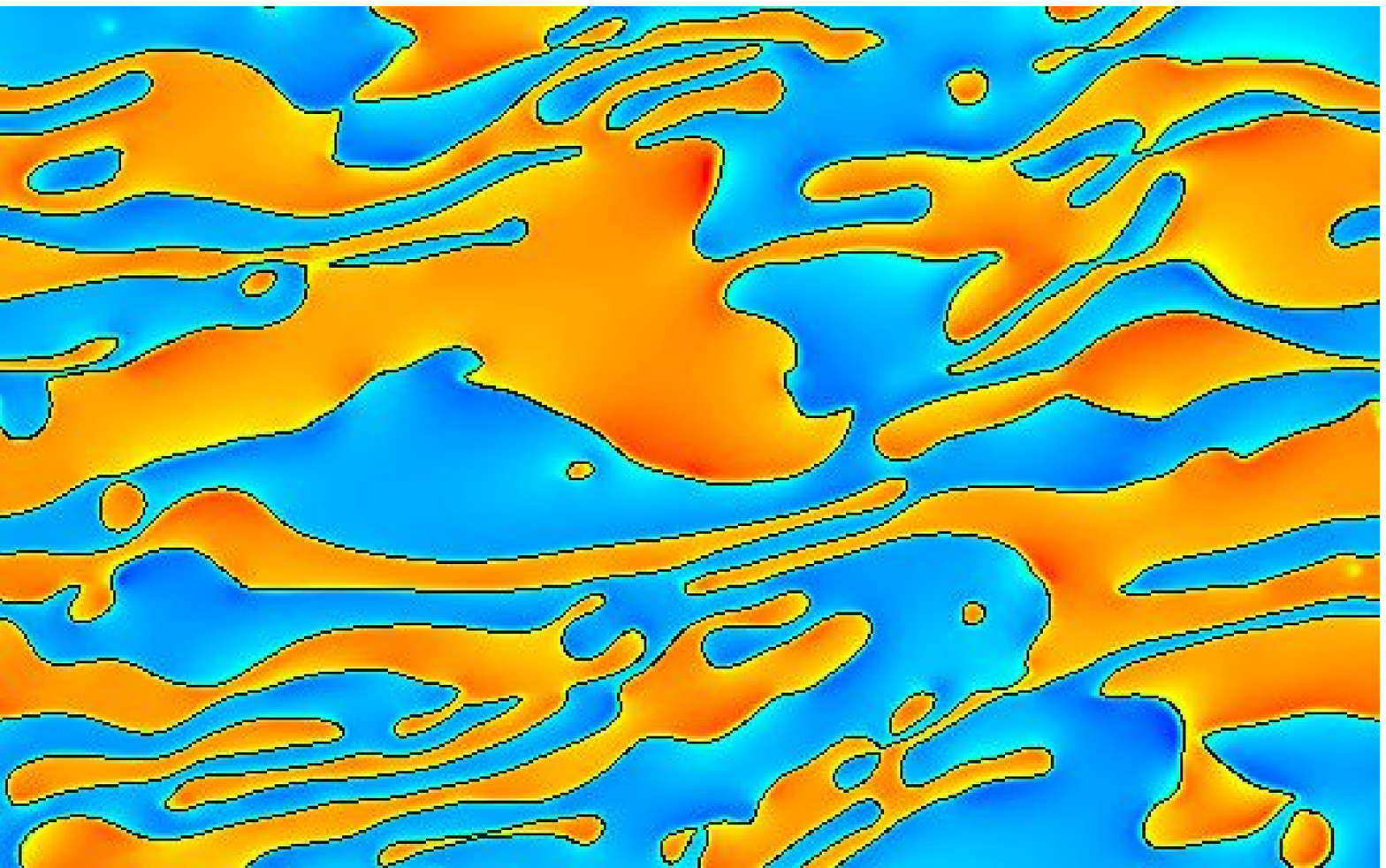}
}
\subfigure[]
{
\includegraphics[width = 0.4\textwidth]{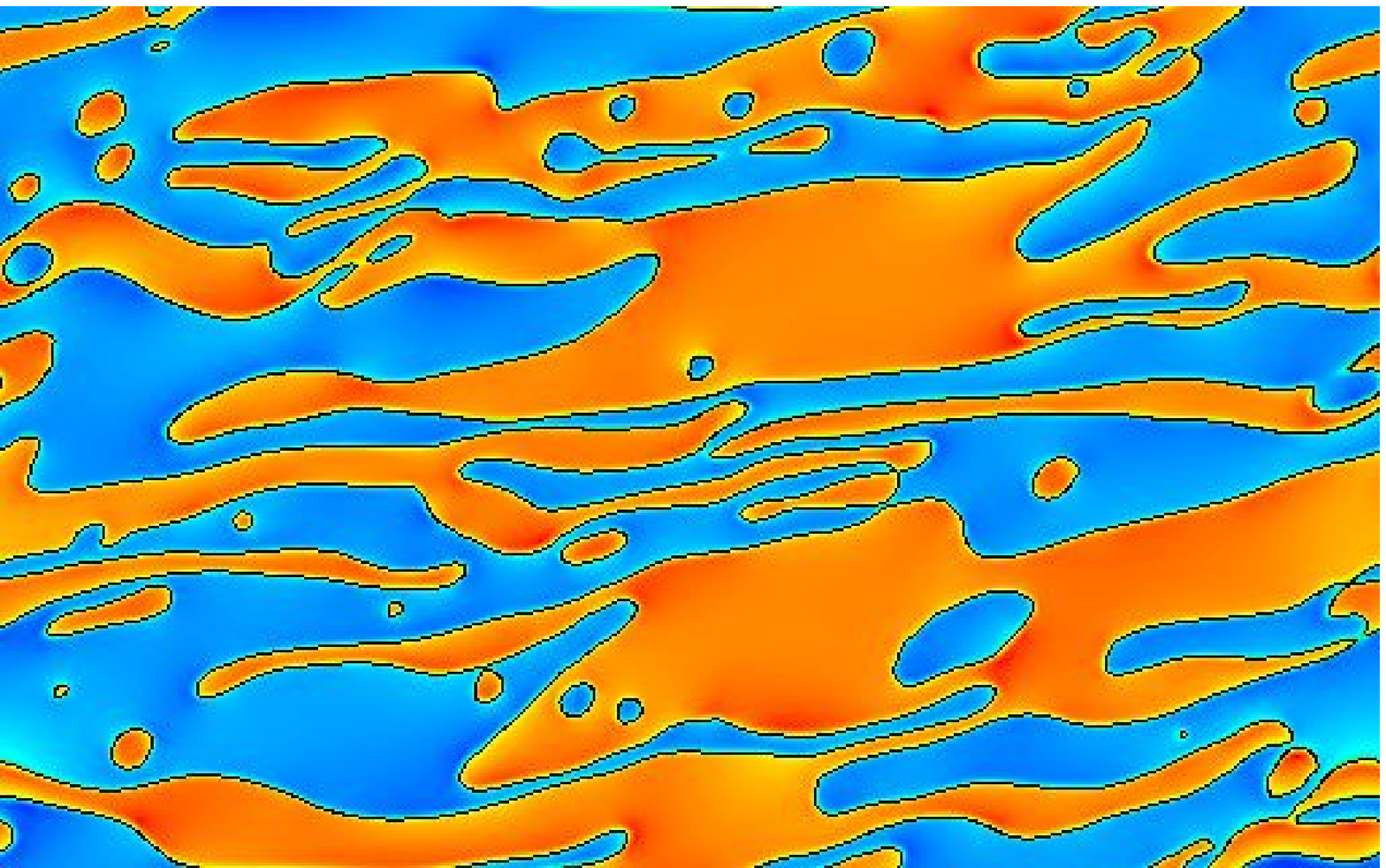} 
}
\caption{Bicontinuous domains at step 90,000 for SPBC with (a) 8 planes and (b) 32 planes \label{spino:SPBCpcolor}. Notice the absence of wrapped domains.}
\end{figure}

\begin{figure} 
\subfigure[]
{
\includegraphics[width = 0.4\textwidth]{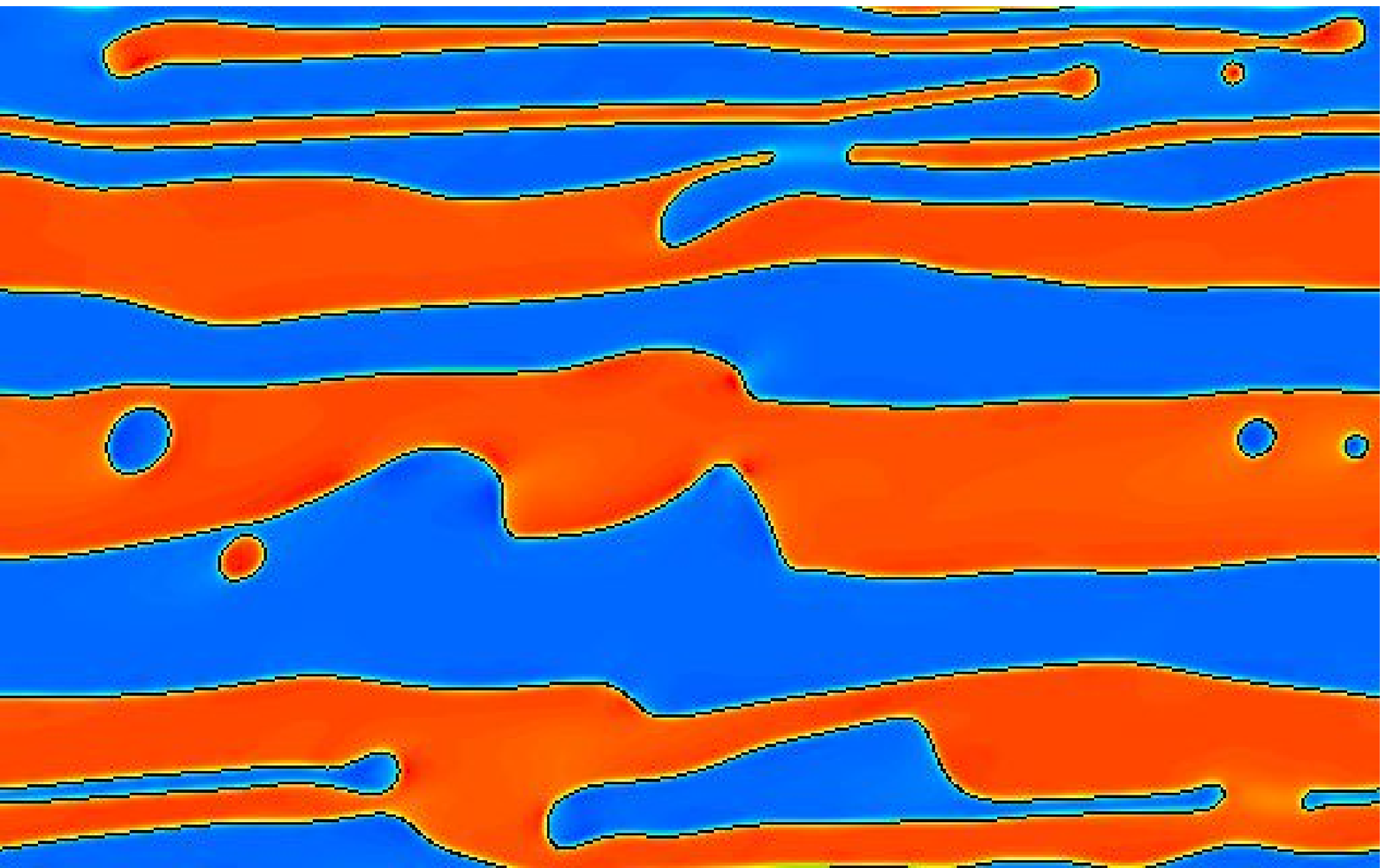}
}
\subfigure[]
{
\includegraphics[width = 0.4\textwidth]{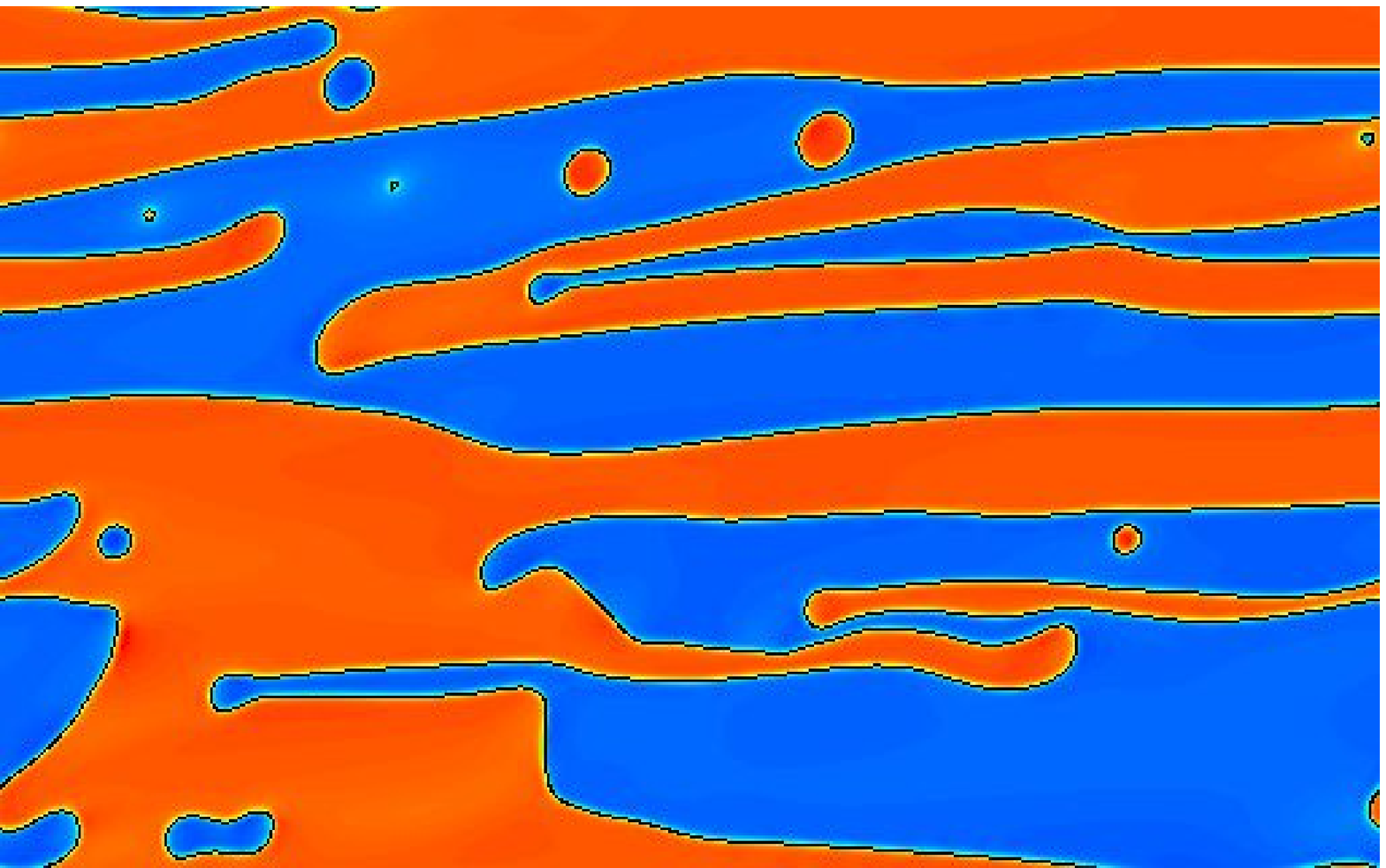} 
}
\caption{Bicontinuous domains at step 90,000 for WP0 with (a) 8 planes and (b) 32 planes. Notice that domains are wrapped. \label{spino:WP0pcolor}}
\end{figure}

\begin{figure} 
\subfigure[]
{
\includegraphics[width = 0.4\textwidth]{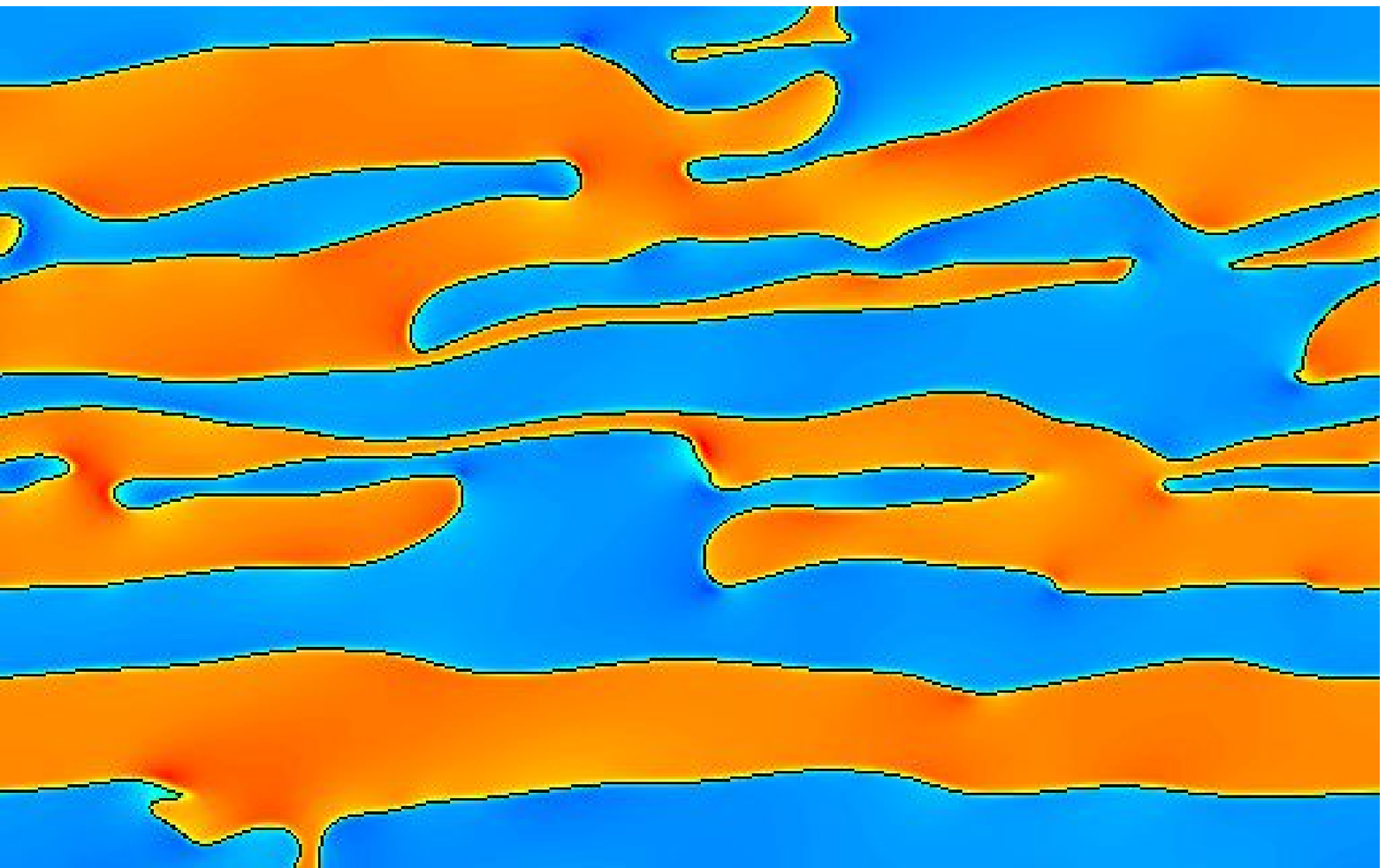}
}
\subfigure[]
{
\includegraphics[width = 0.4\textwidth]{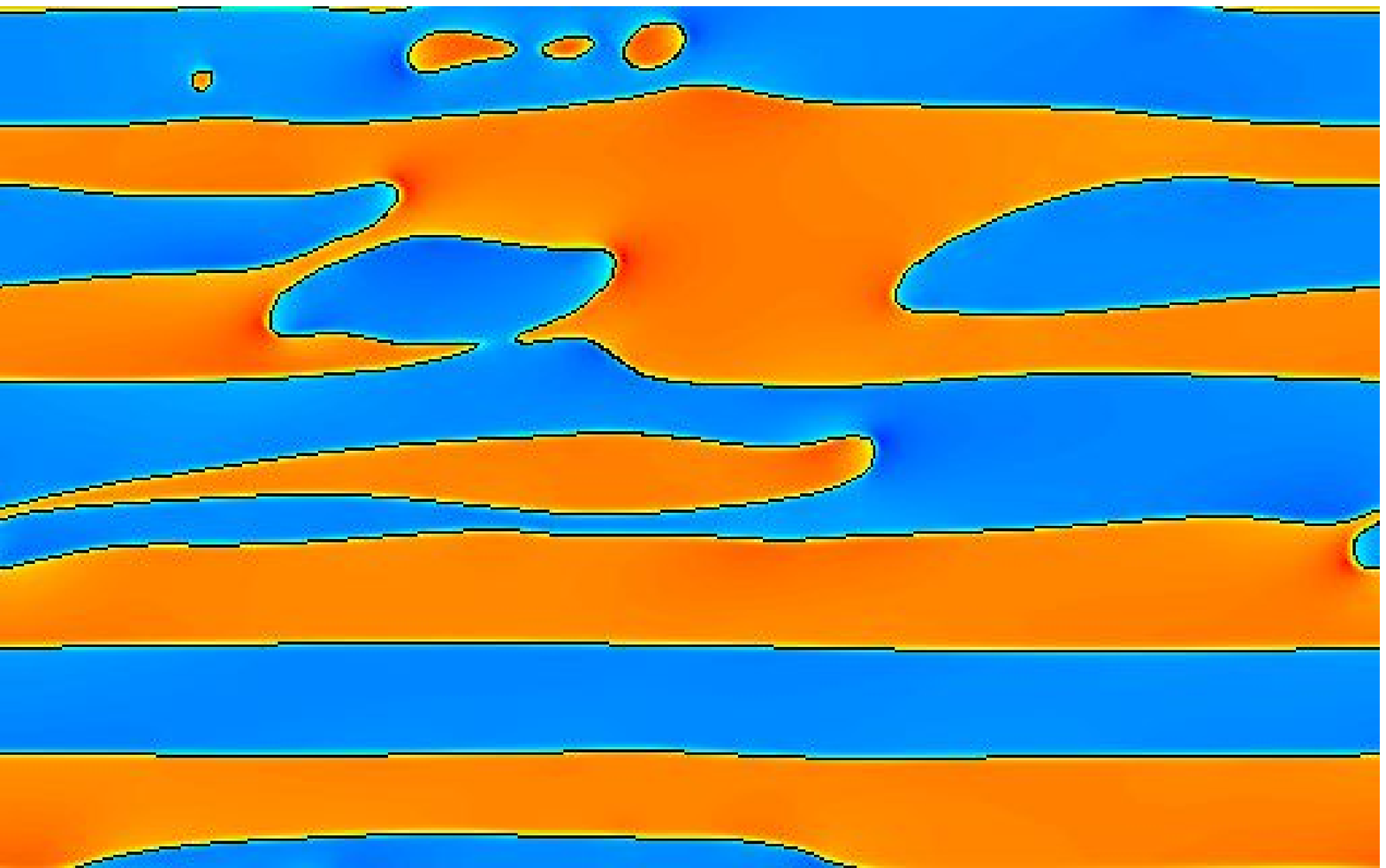} 
}
\caption{Bicontinuous domains at step 90,000 for WP1 with (a) 8 planes and (b) 32 planes. Notice that domains are wrapped.\label{spino:WP1pcolor}}
\end{figure}

\begin{figure} 
\subfigure[]
{
\includegraphics[width = 0.3\textwidth]{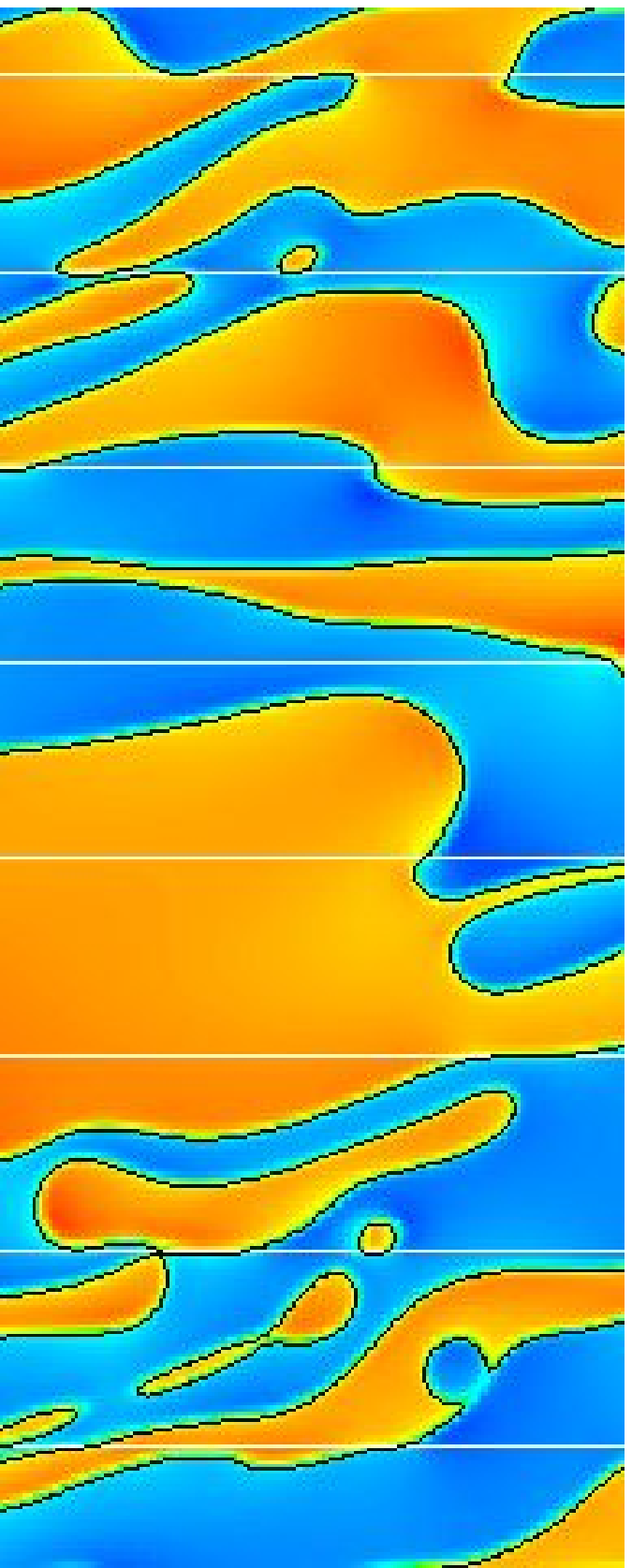}
}
\subfigure[]
{
\includegraphics[width = 0.3\textwidth]{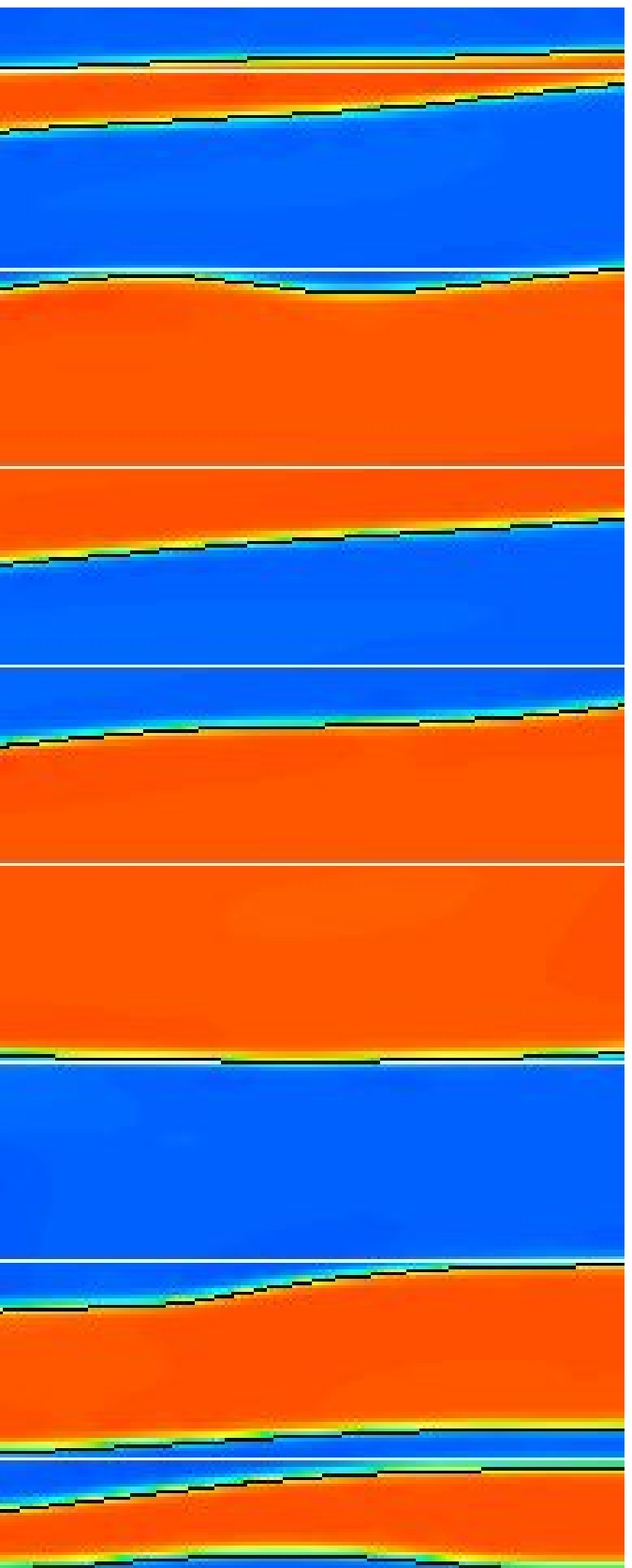} 
}
\subfigure[]
{
\includegraphics[width = 0.3\textwidth]{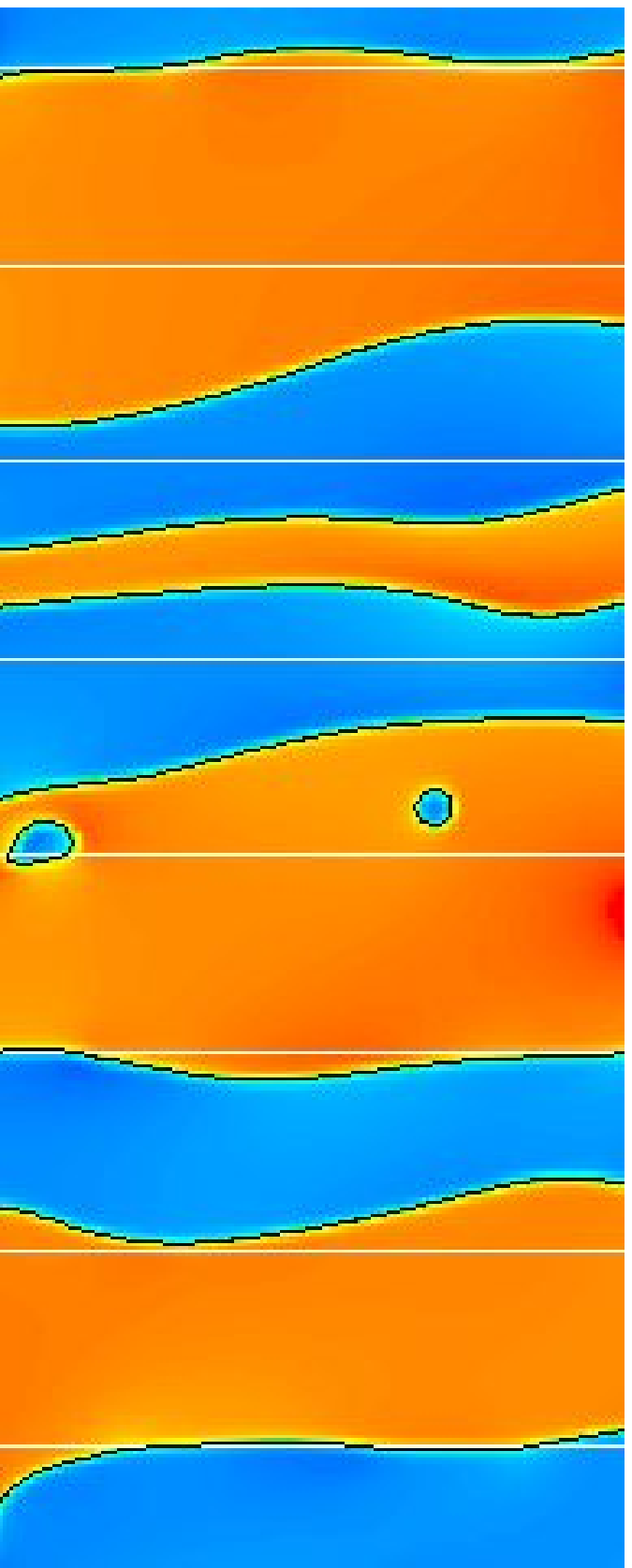} 
}
\caption{Bicontinuous domains: $128 \times 320$ crop at step 80,000
for (a) SPBC, (b) WP0 and (c) WP1. Notice how interfaces are aligned with the LE planes in (b) and (c) but not in (a). More extreme examples of artifacts produced by WP0 can be found in \cite{Cates:2004philtrans}.\label{spino:allcrops}}
\end{figure}

Our next test involves spinodal decomposition in shear flow. We performed two sets of simulations on a 320$\times$512 lattice with 8 and 32 LE planes. In either case, the shear rate was fixed at $\dot{\gamma} = 0.001$, requiring $U=0.04$ for 8 planes and $U=0.01$ for 32 planes. Initial conditions were set for a 50/50 mixture, with a random composition at each site while $f_i$ distributions  were initialised to their steady state values. The remaining parameters were $\eta = 0.025$, $A=-B=0.00625$, $\kappa = 0.004$, $M=4.0$ and $\sigma = 0.0042$. Physically, the results should depend only on $\dot\gamma$ and not on the number of LE planes. Pseudo-colour plots of the domain morphology at late times are shown in Fig.\ref{spino:SPBCpcolor} (SPBC), Fig.\ref{spino:WP0pcolor} (WP0), and Fig.\ref{spino:WP1pcolor} (WP1). The morphology in Fig.\ref{spino:WP0pcolor} and Fig.\ref{spino:WP1pcolor} depends on the number of LE planes. In particular, increasing the number of LE planes increases the number of domains that have ``wrapped'' around the flow direction. We believe this arises due to the error introduced at the LE planes by previous algorithms, as noted in \cite{Cates:2004philtrans}. The nature of this error is to introduce a spurious alignment of the interface with the LE planes, leading to a pinning of the domains and so promoting the wraparound along the flow direction. This pinning effect is  clearly seen in Fig.\ref{spino:allcrops} for the WP0 and WP1 methods, but is not discernable using SPBC.

\begin{figure} 
\subfigure[]
{
\includegraphics[]{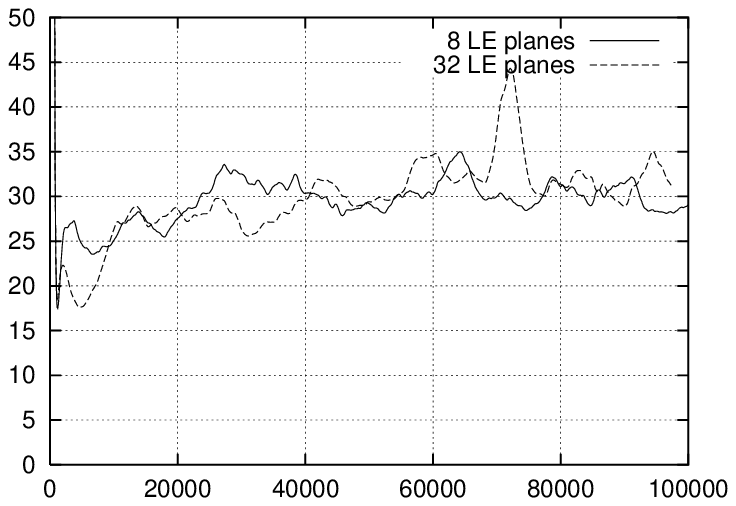}
}
\subfigure[]
{
\includegraphics[]{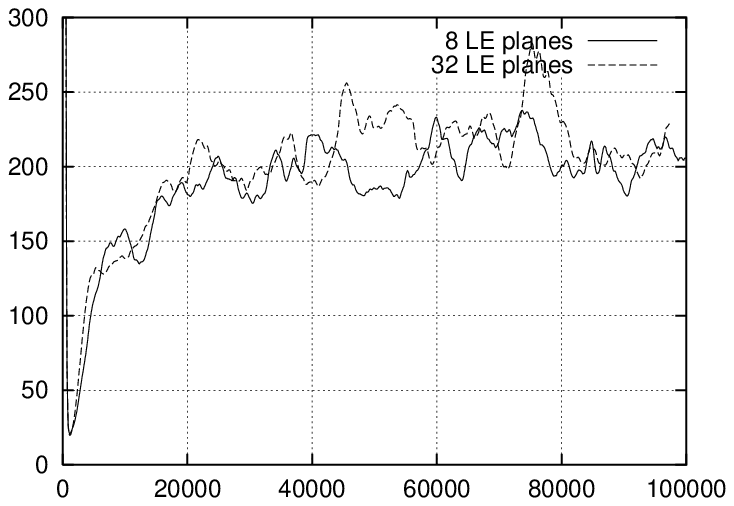} 
}
\caption{Length scales (a) $L1$ and (b) $L2$  for SPBC\label{spino:SPBC}, showing no discernable dependence on the number of planes.}
\end{figure}

\begin{figure} 
\subfigure[]
{
\includegraphics[]{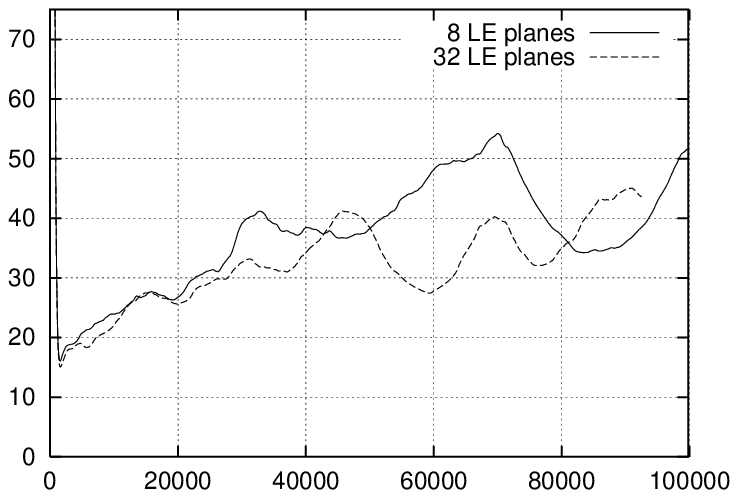}
}
\subfigure[]
{
\includegraphics[]{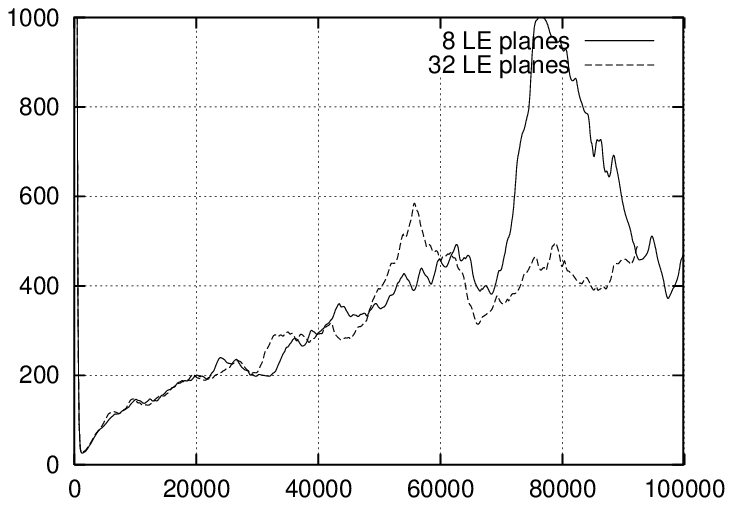} 
}
\caption{Length scales (a) $L_1$ (b) $L_2$ for WP0\label{spino:WP0} showing that $L2$ is weakly dependent on the number of planes.}
\end{figure}

\begin{figure} 
\subfigure[]
{
\includegraphics[]{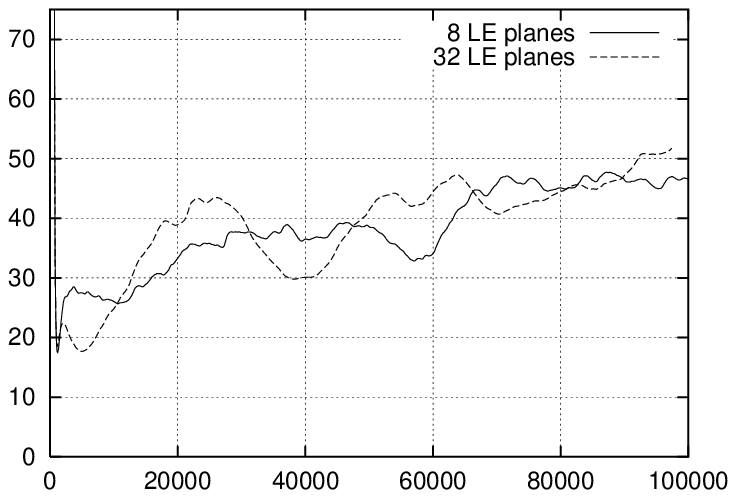}
}
\subfigure[]
{
\includegraphics[]{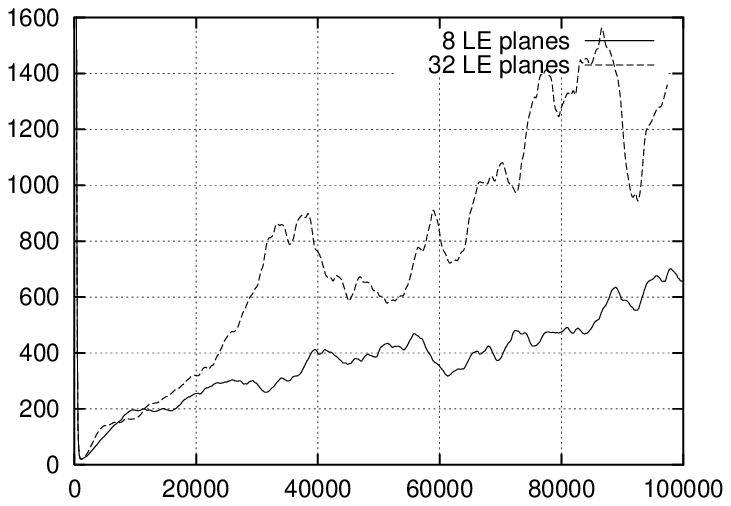} 
}
\caption{Length scales (a) $L_1$ and (b)$L_2$) for WP1\label{spino:WP1} showing that $L2$ depends strongly on the number of planes.}
\end{figure}

A more quantitative test follows from length scale measures $L_1$ and $L_2$ extracted from the gradients of the order parameter field \cite{Wagner:1999, Wagner:2003droplet}. Physically, these denote the principal axes of spherical droplet deformed by the flow, reflecting the effect of elongation (which stretches the droplet into an ellipse whose major axis is aligned with the flow) and rotation (which rotates the principal axes of the ellipse). By convention, $L_1$ is the minor axis and $L_2$ the major axis. Applied to a spinodal pattern, $L_1$ measures the deformation approximately normal to the flow, while $L_2$ measures the deformation approximately in the direction of flow. Simulations in which the domains wrap around the flow direction ( $L_2 > L_x$ ) are contaminated by finite-size effects; no quantitative conclusions can be drawn from such data. In Fig.\ref{spino:SPBC}, Fig.\ref{spino:WP0} and Fig.\ref{spino:WP1} we compare the length scales $L_1$ and $L_2$ for SPBC, WP0, and WP1 respectively. In the latter two cases, $L_2$ gradually increases beyond $L_x$, indicating wrap around, and no steady state is reached. Further, the rate at which the wraparound proceeds (the slope of the $L_2(t)$ curve) increases with the number of LE planes. In the case of SPBC we obtain a steady-state with $L_2 < L_x$, with no dependence on the number of LE planes. This is the first observation, within an LBM simulation using periodic boundary conditions, of a dynamical steady state in which the domain scales $L_1$ and $L_2$ achieve saturation in a manner that is not affected by finite size artifacts and/or lock-in between fluid interfaces and Lees-Edwards planes, as visible in Fig.\ref{spino:allcrops}. More detailed results for this problem will be presented elsewhere \cite{Stansell:2005}.

\subsection{Limitations}
Combining these results, we feel confident that the errors introduced solely due to SPBC and internal LE boundaries are now under good control, and no larger than other sources of error \cite{kendon:2001jfm} in the LBM for binary fluids. This could not be said for the previous state of the art as represented by WP0. However, shear flow may introduce errors beyond those mentioned in \cite{kendon:2001jfm}. The LBM model described here is not Gallilean invariant. This deficiency is likely to show up more acutely in the presence of shear flows, in particular with multiple LE planes. The small artifact at the SPBC planes noted in our droplet test may be due to the lack of Galilean invariance. Our SPBC rules do not break Galilean invariance, however. The interpolation of the distribution functions, while respecting the conservation laws, may introduce numerical errors. We have chosen the simplest linear interpolation scheme, and no doubt more accurate interpolations schemes can be used to improve accuracy. Finally, in shear flow there is always a mean advection velocity, implying that errors introduced by the advection scheme will be accumulative. The advection scheme of the LKE is first order and has a diffusive error which depends on the Courant number $C = U\Delta t/\Delta x$. While a systematic error analysis remains to be done, we have noted that acceptable accuracy is obtained only with $C < 0.05$. For larger values, the simulation is contaminated by the diffusive error of the advection scheme. 

\section{Conclusion}
To summarise, we have presented a new scheme for the shear flow in lattice Boltzmann and lattice kinetic models, which follows from a systematic coarse-graining of the Lees-Edwards boundary conditions of molecular dynamics. The boundary conditions are then implemented on the discrete kinetic space. The problem of discrete velocities is solved by applying the transformation to the moments of the distribution functions, while the problem of discrete space is solved by a numerical interpolation. We have validated our method for one-component and two-component fluids and find excellent agreement with analytical results when available, and qualitatively correct behaviour in more complicated cases. The present algorithm allows for the investigation of the effects of shear on a wide variety of complex fluid flows. The present method may be combined with our fluctuating lattice Boltzmann equation (FLBE) \cite{Adhikari:2004} to study the interplay of thermal fluctuations and shear in complex fluid hydrodynamics. We are presently investigating the effects of shear on spinodal decomposition in binary fluids \cite{Stansell:2005}.

\begin{acknowledgements}
We thank Ignacio Pagonabarraga, and Paul Stansell for useful discussions and Mike Cates for helpful suggestions and a critical reading of the manuscript. R. A. and K. S. are supported by EPSRC GR/R67699 (Reality Grid).  
\end{acknowledgements}

\bibliographystyle{apsrev}
\bibliography{leesedwards}

\begin{thebibliography}{41}
\expandafter\ifx\csname natexlab\endcsname\relax\def\natexlab#1{#1}\fi
\expandafter\ifx\csname bibnamefont\endcsname\relax
  \def\bibnamefont#1{#1}\fi
\expandafter\ifx\csname bibfnamefont\endcsname\relax
  \def\bibfnamefont#1{#1}\fi
\expandafter\ifx\csname citenamefont\endcsname\relax
  \def\citenamefont#1{#1}\fi
\expandafter\ifx\csname url\endcsname\relax
  \def\url#1{\texttt{#1}}\fi
\expandafter\ifx\csname urlprefix\endcsname\relax\def\urlprefix{URL }\fi
\providecommand{\bibinfo}[2]{#2}
\providecommand{\eprint}[2][]{\url{#2}}

\bibitem[{\citenamefont{Cates and Evans}(2000)}]{Cates:2000asi}
\bibinfo{editor}{\bibfnamefont{M.~E.} \bibnamefont{Cates}} \bibnamefont{and}
  \bibinfo{editor}{\bibfnamefont{M.~R.} \bibnamefont{Evans}}, eds.,
  \emph{\bibinfo{title}{Soft and {F}ragile {M}atter: {N}onequilibrium
  {D}ynamics, {M}etastability and {F}low}}, no.~\bibinfo{number}{53} in
  \bibinfo{series}{Scottish Graduate Textbook Series}
  (\bibinfo{publisher}{Institute of Physics}, \bibinfo{year}{2000}).

\bibitem[{\citenamefont{Onuki}(1997)}]{Onuki:1997}
\bibinfo{author}{\bibfnamefont{A.}~\bibnamefont{Onuki}}, \bibinfo{journal}{J.
  Phys. Cond. Matt} \textbf{\bibinfo{volume}{9}}, \bibinfo{pages}{6119}
  (\bibinfo{year}{1997}).

\bibitem[{\citenamefont{Ackerson and Clark}(1981)}]{Ackerson:1981}
\bibinfo{author}{\bibfnamefont{B.~J.} \bibnamefont{Ackerson}} \bibnamefont{and}
  \bibinfo{author}{\bibfnamefont{N.~A.} \bibnamefont{Clark}},
  \bibinfo{journal}{Phys. Rev. Lett} \textbf{\bibinfo{volume}{46}},
  \bibinfo{pages}{123} (\bibinfo{year}{1981}).

\bibitem[{\citenamefont{Cates and Milner}(1989)}]{Cates:1989i2l}
\bibinfo{author}{\bibfnamefont{M.~E.} \bibnamefont{Cates}} \bibnamefont{and}
  \bibinfo{author}{\bibfnamefont{S.~T.} \bibnamefont{Milner}},
  \bibinfo{journal}{Phys. Rev. Lett} \textbf{\bibinfo{volume}{62}},
  \bibinfo{pages}{1856} (\bibinfo{year}{1989}).

\bibitem[{\citenamefont{Bandyopadhyay et~al.}(2000)\citenamefont{Bandyopadhyay,
  Basappa, and Sood}}]{Bandyopadhyay:2000}
\bibinfo{author}{\bibfnamefont{R.}~\bibnamefont{Bandyopadhyay}},
  \bibinfo{author}{\bibfnamefont{G.}~\bibnamefont{Basappa}}, \bibnamefont{and}
  \bibinfo{author}{\bibfnamefont{A.~K.} \bibnamefont{Sood}},
  \bibinfo{journal}{Phys. Rev. Lett} \textbf{\bibinfo{volume}{84}},
  \bibinfo{pages}{2022} (\bibinfo{year}{2000}).

\bibitem[{\citenamefont{Lees and Edwards}(1975)}]{Lees:1972}
\bibinfo{author}{\bibfnamefont{A.}~\bibnamefont{Lees}} \bibnamefont{and}
  \bibinfo{author}{\bibfnamefont{S.~F.} \bibnamefont{Edwards}},
  \bibinfo{journal}{J Phys. C} \textbf{\bibinfo{volume}{5}},
  \bibinfo{pages}{1921} (\bibinfo{year}{1975}).

\bibitem[{\citenamefont{Evans and Morris}(1990)}]{Evans:1990book}
\bibinfo{author}{\bibnamefont{Evans}} \bibnamefont{and}
  \bibinfo{author}{\bibnamefont{Morris}}, \emph{\bibinfo{title}{Statistical
  {M}echanics of {N}onequilibrium {L}iquids}} (\bibinfo{publisher}{Academic
  Press, London}, \bibinfo{year}{1990}), \bibinfo{note}{available online at
  http://rsc.anu.edu.au/~evans/evansmorrissbook.htm}.

\bibitem[{\citenamefont{Liboff}(1990)}]{Liboff:1990book}
\bibinfo{author}{\bibfnamefont{R.~L.} \bibnamefont{Liboff}},
  \emph{\bibinfo{title}{Kinetic {T}heory}} (\bibinfo{publisher}{Prentice Hall,
  Englewood Cliffs, New Jersey}, \bibinfo{year}{1990}).

\bibitem[{\citenamefont{Chapman and Cowling}(1991)}]{Chapman:1991book}
\bibinfo{author}{\bibfnamefont{S.}~\bibnamefont{Chapman}} \bibnamefont{and}
  \bibinfo{author}{\bibfnamefont{T.~G.} \bibnamefont{Cowling}},
  \emph{\bibinfo{title}{The {M}athematical {T}heory of {N}on-{U}niform
  {G}ases}} (\bibinfo{publisher}{Cambridge University Press, Cambridge},
  \bibinfo{year}{1991}), \bibinfo{edition}{3rd} ed.

\bibitem[{\citenamefont{Wagner and Yeomans}(1999)}]{Wagner:1999}
\bibinfo{author}{\bibfnamefont{A.~J.} \bibnamefont{Wagner}} \bibnamefont{and}
  \bibinfo{author}{\bibfnamefont{J.~M.} \bibnamefont{Yeomans}},
  \bibinfo{journal}{Phys. Rev. E} \textbf{\bibinfo{volume}{59}},
  \bibinfo{pages}{4366} (\bibinfo{year}{1999}).

\bibitem[{\citenamefont{Wagner and Pagonabarraga}(2002)}]{Wagner:2002}
\bibinfo{author}{\bibfnamefont{A.~J.} \bibnamefont{Wagner}} \bibnamefont{and}
  \bibinfo{author}{\bibfnamefont{I.}~\bibnamefont{Pagonabarraga}},
  \bibinfo{journal}{J. Stat. Phys} \textbf{\bibinfo{volume}{107}},
  \bibinfo{pages}{521} (\bibinfo{year}{2002}).

\bibitem[{\citenamefont{Chen and Doolen}(1998)}]{chen:98}
\bibinfo{author}{\bibfnamefont{S.}~\bibnamefont{Chen}} \bibnamefont{and}
  \bibinfo{author}{\bibfnamefont{G.}~\bibnamefont{Doolen}},
  \bibinfo{journal}{Annu. Rev. Fluid Mech.} \textbf{\bibinfo{volume}{30}},
  \bibinfo{pages}{329} (\bibinfo{year}{1998}).

\bibitem[{\citenamefont{Succi}(2001)}]{succi-book:01}
\bibinfo{author}{\bibfnamefont{S.}~\bibnamefont{Succi}},
  \emph{\bibinfo{title}{The {L}attice {B}oltzmann {E}quation for {F}luid
  {D}ynamics and {B}eyond}} (\bibinfo{publisher}{Clarendon Press, Oxford},
  \bibinfo{year}{2001}).

\bibitem[{\citenamefont{McNamara and Zanetti}(1988)}]{McNamara:88}
\bibinfo{author}{\bibfnamefont{G.}~\bibnamefont{McNamara}} \bibnamefont{and}
  \bibinfo{author}{\bibfnamefont{G.}~\bibnamefont{Zanetti}},
  \bibinfo{journal}{Phys. Rev. Lett.} \textbf{\bibinfo{volume}{61}},
  \bibinfo{pages}{2332} (\bibinfo{year}{1988}).

\bibitem[{\citenamefont{Higuera et~al.}(1989)\citenamefont{Higuera, Jimenez,
  and Succi}}]{higuera:89}
\bibinfo{author}{\bibfnamefont{F.}~\bibnamefont{Higuera}},
  \bibinfo{author}{\bibfnamefont{J.}~\bibnamefont{Jimenez}}, \bibnamefont{and}
  \bibinfo{author}{\bibfnamefont{S.}~\bibnamefont{Succi}},
  \bibinfo{journal}{Europhys. Lett} \textbf{\bibinfo{volume}{9}},
  \bibinfo{pages}{663} (\bibinfo{year}{1989}).

\bibitem[{\citenamefont{Shan and He}(1998)}]{Shan:1998}
\bibinfo{author}{\bibfnamefont{X.}~\bibnamefont{Shan}} \bibnamefont{and}
  \bibinfo{author}{\bibfnamefont{X.}~\bibnamefont{He}}, \bibinfo{journal}{Phys.
  Rev. Lett} \textbf{\bibinfo{volume}{80}}, \bibinfo{pages}{65}
  (\bibinfo{year}{1998}).

\bibitem[{\citenamefont{He and Luo}(1997)}]{He:1997b}
\bibinfo{author}{\bibfnamefont{X.}~\bibnamefont{He}} \bibnamefont{and}
  \bibinfo{author}{\bibfnamefont{L.-S.} \bibnamefont{Luo}},
  \bibinfo{journal}{Phys. Rev. E} \textbf{\bibinfo{volume}{55}},
  \bibinfo{pages}{R6333} (\bibinfo{year}{1997}).

\bibitem[{\citenamefont{Perthame}(2002)}]{Perthame:2002book}
\bibinfo{author}{\bibfnamefont{B.}~\bibnamefont{Perthame}},
  \emph{\bibinfo{title}{Kinetic {F}ormulation of {C}onservation {L}aws}}
  (\bibinfo{publisher}{Clarendon Press, Oxford}, \bibinfo{year}{2002}).

\bibitem[{\citenamefont{Swift et~al.}(1996)\citenamefont{Swift, Olrandini,
  Osborn, and Yeomans}}]{Swift:96}
\bibinfo{author}{\bibfnamefont{M.}~\bibnamefont{Swift}},
  \bibinfo{author}{\bibfnamefont{E.}~\bibnamefont{Olrandini}},
  \bibinfo{author}{\bibfnamefont{W.}~\bibnamefont{Osborn}}, \bibnamefont{and}
  \bibinfo{author}{\bibfnamefont{J.}~\bibnamefont{Yeomans}},
  \bibinfo{journal}{Phys. Rev. E} \textbf{\bibinfo{volume}{54}},
  \bibinfo{pages}{5051} (\bibinfo{year}{1996}).

\bibitem[{\citenamefont{Koelman}(1991)}]{Koelman:91}
\bibinfo{author}{\bibfnamefont{J.}~\bibnamefont{Koelman}},
  \bibinfo{journal}{Europhysics Lett.} \textbf{\bibinfo{volume}{15}},
  \bibinfo{pages}{603} (\bibinfo{year}{1991}).

\bibitem[{\citenamefont{Qian et~al.}(1992)\citenamefont{Qian, d'Humieres, and
  Lallemand}}]{Qian:92}
\bibinfo{author}{\bibfnamefont{Y.}~\bibnamefont{Qian}},
  \bibinfo{author}{\bibfnamefont{D.}~\bibnamefont{d'Humieres}},
  \bibnamefont{and}
  \bibinfo{author}{\bibfnamefont{P.}~\bibnamefont{Lallemand}},
  \bibinfo{journal}{Europhysics Lett.} \textbf{\bibinfo{volume}{17}},
  \bibinfo{pages}{479} (\bibinfo{year}{1992}).

\bibitem[{\citenamefont{Ladd and Verberg}(2001)}]{Ladd:2001}
\bibinfo{author}{\bibfnamefont{A.~J.~C.} \bibnamefont{Ladd}} \bibnamefont{and}
  \bibinfo{author}{\bibfnamefont{R.}~\bibnamefont{Verberg}},
  \bibinfo{journal}{J. Stat. Phys} \textbf{\bibinfo{volume}{104}},
  \bibinfo{pages}{1191} (\bibinfo{year}{2001}).

\bibitem[{\citenamefont{He et~al.}(1998)\citenamefont{He, Shan, and
  Doolen}}]{He:1998nonideal}
\bibinfo{author}{\bibfnamefont{X.}~\bibnamefont{He}},
  \bibinfo{author}{\bibfnamefont{X.}~\bibnamefont{Shan}}, \bibnamefont{and}
  \bibinfo{author}{\bibfnamefont{G.~D.} \bibnamefont{Doolen}},
  \bibinfo{journal}{Phys. Rev. E.} \textbf{\bibinfo{volume}{57}},
  \bibinfo{pages}{R13 } (\bibinfo{year}{1998}).

\bibitem[{\citenamefont{Dellar}(2003)}]{Dellar:2003}
\bibinfo{author}{\bibfnamefont{P.}~\bibnamefont{Dellar}}, \bibinfo{journal}{J.
  Comp. Phys} \textbf{\bibinfo{volume}{190}}, \bibinfo{pages}{351}
  (\bibinfo{year}{2003}).

\bibitem[{\citenamefont{Chaikin and Lubensky}(2000)}]{Chaikin:2000}
\bibinfo{author}{\bibfnamefont{P.}~\bibnamefont{Chaikin}} \bibnamefont{and}
  \bibinfo{author}{\bibfnamefont{T.}~\bibnamefont{Lubensky}},
  \emph{\bibinfo{title}{Principles of {C}ondensed {M}atter {P}hysics}}
  (\bibinfo{publisher}{Cambridge University Press, Cambridge},
  \bibinfo{year}{2000}).

\bibitem[{\citenamefont{Stratford et~al.}(2005)\citenamefont{Stratford,
  Adhikari, Pagonabarraga, and Desplat}}]{Stratford:2005}
\bibinfo{author}{\bibfnamefont{K.}~\bibnamefont{Stratford}},
  \bibinfo{author}{\bibfnamefont{R.}~\bibnamefont{Adhikari}},
  \bibinfo{author}{\bibfnamefont{I.}~\bibnamefont{Pagonabarraga}},
  \bibnamefont{and} \bibinfo{author}{\bibfnamefont{J.~C.}
  \bibnamefont{Desplat}} (\bibinfo{year}{2005}), \bibinfo{note}{submitted to J.
  Stat. Phys, e-print cond-mat/0407631}.

\bibitem[{\citenamefont{Kendon et~al.}(2001)\citenamefont{Kendon, Cates,
  Pagonabarraga, Desplat, and Bladon}}]{kendon:2001jfm}
\bibinfo{author}{\bibfnamefont{V.~M.} \bibnamefont{Kendon}},
  \bibinfo{author}{\bibfnamefont{M.~E.} \bibnamefont{Cates}},
  \bibinfo{author}{\bibfnamefont{I.}~\bibnamefont{Pagonabarraga}},
  \bibinfo{author}{\bibfnamefont{J.~C.} \bibnamefont{Desplat}},
  \bibnamefont{and} \bibinfo{author}{\bibfnamefont{P.}~\bibnamefont{Bladon}},
  \bibinfo{journal}{J. Fluid. Mech} \textbf{\bibinfo{volume}{440}},
  \bibinfo{pages}{147} (\bibinfo{year}{2001}).

\bibitem[{\citenamefont{Luo}(2000)}]{Luo:2000}
\bibinfo{author}{\bibfnamefont{L.~S.} \bibnamefont{Luo}},
  \bibinfo{journal}{Phys. Rev. E} \textbf{\bibinfo{volume}{62}},
  \bibinfo{pages}{4982} (\bibinfo{year}{2000}).

\bibitem[{\citenamefont{Adhikari et~al.}(2005)\citenamefont{Adhikari,
  Pagonabarraga, and Stratford}}]{Adhikari:2005}
\bibinfo{author}{\bibfnamefont{R.}~\bibnamefont{Adhikari}},
  \bibinfo{author}{\bibfnamefont{I.}~\bibnamefont{Pagonabarraga}},
  \bibnamefont{and} \bibinfo{author}{\bibfnamefont{K.}~\bibnamefont{Stratford}}
  (\bibinfo{year}{2005}), \bibinfo{note}{in preparation}.

\bibitem[{\citenamefont{Cercignani}(1988)}]{Cercignani:1988book}
\bibinfo{author}{\bibfnamefont{C.}~\bibnamefont{Cercignani}},
  \emph{\bibinfo{title}{The {B}oltzmann {E}quation and its {A}pplications}}
  (\bibinfo{publisher}{Springer-Verlag, New York}, \bibinfo{year}{1988}).

\bibitem[{\citenamefont{Ansumali and Karlin}(2002)}]{Ansumali:2002}
\bibinfo{author}{\bibfnamefont{S.}~\bibnamefont{Ansumali}} \bibnamefont{and}
  \bibinfo{author}{\bibfnamefont{I.}~\bibnamefont{Karlin}},
  \bibinfo{journal}{Phys. Rev. E} \textbf{\bibinfo{volume}{66}},
  \bibinfo{pages}{026311} (\bibinfo{year}{2002}).

\bibitem[{\citenamefont{Grad}(1949)}]{Grad:1949}
\bibinfo{author}{\bibfnamefont{H.}~\bibnamefont{Grad}},
  \bibinfo{journal}{Commun. Pure Appl. Math} \textbf{\bibinfo{volume}{2}},
  \bibinfo{pages}{325} (\bibinfo{year}{1949}).

\bibitem[{\citenamefont{Benzi et~al.}(1992)\citenamefont{Benzi, Succi, and
  Vergassola}}]{benzi:92}
\bibinfo{author}{\bibfnamefont{R.}~\bibnamefont{Benzi}},
  \bibinfo{author}{\bibfnamefont{S.}~\bibnamefont{Succi}}, \bibnamefont{and}
  \bibinfo{author}{\bibfnamefont{M.}~\bibnamefont{Vergassola}},
  \bibinfo{journal}{Physics Reports} \textbf{\bibinfo{volume}{222}},
  \bibinfo{pages}{145} (\bibinfo{year}{1992}).

\bibitem[{\citenamefont{Dellar}(2002)}]{Dellar:2002}
\bibinfo{author}{\bibfnamefont{P.}~\bibnamefont{Dellar}},
  \bibinfo{journal}{Phys. Rev. E} \textbf{\bibinfo{volume}{65}},
  \bibinfo{pages}{036309} (\bibinfo{year}{2002}).

\bibitem[{\citenamefont{d'Humieres}(1992)}]{dHumieres:1992}
\bibinfo{author}{\bibfnamefont{D.}~\bibnamefont{d'Humieres}}, in
  \emph{\bibinfo{booktitle}{Rarefied {G}as {D}ynamics: {T}heory and
  {S}imulations}}, edited by \bibinfo{editor}{\bibfnamefont{B.~D.}
  \bibnamefont{Shizgal}} \bibnamefont{and}
  \bibinfo{editor}{\bibfnamefont{D.~P.} \bibnamefont{Weaver}}
  (\bibinfo{organization}{AIAA, Washington, D.C.}, \bibinfo{year}{1992}), vol.
  \bibinfo{volume}{159} of \emph{\bibinfo{series}{Prog. Astronaut. Aeronaut}},
  pp. \bibinfo{pages}{450--458}.

\bibitem[{\citenamefont{Adhikari et~al.}(2004)\citenamefont{Adhikari, Cates,
  Stratford, and Wagner}}]{Adhikari:2004}
\bibinfo{author}{\bibfnamefont{R.}~\bibnamefont{Adhikari}},
  \bibinfo{author}{\bibfnamefont{M.~E.} \bibnamefont{Cates}},
  \bibinfo{author}{\bibfnamefont{K.}~\bibnamefont{Stratford}},
  \bibnamefont{and} \bibinfo{author}{\bibfnamefont{A.~J.} \bibnamefont{Wagner}}
  (\bibinfo{year}{2004}), \bibinfo{note}{e-print cond-mat/0402598}.

\bibitem[{\citenamefont{Behrend et~al.}(1994)\citenamefont{Behrend, Harris, and
  Warren}}]{Behrend:1994}
\bibinfo{author}{\bibfnamefont{O.}~\bibnamefont{Behrend}},
  \bibinfo{author}{\bibfnamefont{R.}~\bibnamefont{Harris}}, \bibnamefont{and}
  \bibinfo{author}{\bibfnamefont{P.~B.} \bibnamefont{Warren}},
  \bibinfo{journal}{Phys. Rev. E} \textbf{\bibinfo{volume}{50}},
  \bibinfo{pages}{4586} (\bibinfo{year}{1994}).

\bibitem[{\citenamefont{Batchelor}(1967)}]{Batchelor:1967}
\bibinfo{author}{\bibfnamefont{G.~K.} \bibnamefont{Batchelor}},
  \emph{\bibinfo{title}{An {I}ntroduction to {F}luid {D}ynamics}}
  (\bibinfo{publisher}{Cambridge University Press, Cambridge},
  \bibinfo{year}{1967}).

\bibitem[{\citenamefont{Cates et~al.}(2004)\citenamefont{Cates, Desplat,
  Stansell, Wagner, Stratford, and Pagonabarraga}}]{Cates:2004philtrans}
\bibinfo{author}{\bibfnamefont{M.~E.} \bibnamefont{Cates}},
  \bibinfo{author}{\bibfnamefont{J.~C.} \bibnamefont{Desplat}},
  \bibinfo{author}{\bibfnamefont{P.}~\bibnamefont{Stansell}},
  \bibinfo{author}{\bibfnamefont{A.~J.} \bibnamefont{Wagner}},
  \bibinfo{author}{\bibfnamefont{K.}~\bibnamefont{Stratford}},
  \bibnamefont{and}
  \bibinfo{author}{\bibfnamefont{I.}~\bibnamefont{Pagonabarraga}}
  (\bibinfo{year}{2004}), \bibinfo{note}{e-print cond-mat/0411490}.

\bibitem[{\citenamefont{Wagner et~al.}(2003)\citenamefont{Wagner, WIlson, and
  Cates}}]{Wagner:2003droplet}
\bibinfo{author}{\bibfnamefont{A.~J.} \bibnamefont{Wagner}},
  \bibinfo{author}{\bibfnamefont{L.~M.} \bibnamefont{WIlson}},
  \bibnamefont{and} \bibinfo{author}{\bibfnamefont{M.~E.} \bibnamefont{Cates}},
  \bibinfo{journal}{Phys. Rev. E} \textbf{\bibinfo{volume}{68}},
  \bibinfo{pages}{045301} (\bibinfo{year}{2003}).

\bibitem[{\citenamefont{Stansell~et al}(2005)}]{Stansell:2005}
\bibinfo{author}{\bibfnamefont{P.}~\bibnamefont{Stansell~et al}}
  (\bibinfo{year}{2005}), \bibinfo{note}{in preparation}.

\end{thebibliography}



\begin{table}[h]
\begin{center}
\begin{tabular}{|l|r|r|r|r|r|}
\hline
$t_{\nu}$ & \multicolumn{5}{c|}{$\eta$}\\
        &     1.41&       0.5&       0.2&   0.0065&  0.0005\\
\hline
0.00100 & 4.07e-7& -1.42e-4& -1.10e-3& 2.33e-2& 2.45e-2\\
0.00316 & 5.05e-6& -2.57e-5& -1.89e-4& 4.15e-3& 4.36e-3\\
0.01000 &    0.00& -5.07e-6& -3.13e-5& 7.18e-4& 7.62e-4\\
0.03160 &    0.00& -5.04e-6& -5.12e-6& 1.26e-4& 1.37e-4\\
0.10000 &    0.00&     0.00&     0.00& 1.55e-5& 2.07e-5\\
\hline
\end{tabular}
\caption{Relative error in the velocity $\epsilon_v$ (defined in the text) as a function of dimensionless time $t_{\nu} = \eta t/L_y^2$, half a lattice site from the SPBC plane for different values of the viscosity $\eta$.\label{tab:sf1-all-relerr-LEMRT}}
\end{center}
\end{table}
\end{document}